\documentclass[sigconf]{acmart} 

\renewcommand\footnotetextcopyrightpermission[1]{} 
\acmConference[KnOD'21 Workshop]{}{April 14, 2021}{Virtual Event}

\begin{document}

\title[REMOD]{REMOD: Relation Extraction for Modeling Online Discourse}

\author{Matthew Sumpter}
\email{mjsumpter@usf.edu}
\affiliation{%
  \institution{University of South Florida}
}

\author{Giovanni Luca Ciampaglia}
\email{glc3@mail.usf.edu}
\affiliation{%
  \institution{University of South Florida}
}

\begin{abstract}
The enormous amount of discourse taking place online poses challenges to the functioning of a civil and informed public sphere.
Efforts to standardize online discourse data, such as ClaimReview, are making available a wealth of new data about potentially inaccurate claims, reviewed by third-party fact-checkers. These data could help shed light on the nature of online discourse, the role of political elites in amplifying it, and its implications for the integrity of the online information ecosystem.
Unfortunately, the semi-structured nature of much of this data presents significant challenges when it comes to modeling and reasoning about online discourse.
A key challenge is relation extraction, which is the task of determining the semantic relationships between named entities in a claim.
Here we develop a novel supervised learning method for relation extraction that combines graph embedding techniques with path traversal on semantic dependency graphs. 
Our approach is based on the intuitive observation that knowledge of the entities along the path between the subject and object of a triple (e.g.~\texttt{Washington,\_D.C.}, and \texttt{United\_States\_of\_America}) provides useful information that can be leveraged for extracting its semantic relation (i.e.~\texttt{capitalOf}).
As an example of a potential application of this technique for modeling online discourse, we show that our method can be integrated into a pipeline to reason about potential misinformation claims.
\end{abstract}



\maketitle

\section{Introduction}

The prevalence of false and inaccurate information in its myriad of forms --- a persistent and dangerous societal problem --- is still a poorly understood phenomenon~\cite{allcott2017social, lazer2018science, bovet2019influence}, especially in the context of political communication~\cite{guess2020exposure}. Even though strong exposure to so-called ``fake news'' is limited to the segment of most active news consumers~\cite{grinberg2019fake}, individual claims echoing the false or misleading content shared by these audiences can spread rapidly through social media~\cite{Vosoughi1146, zhao2020fake}, amplified by bots~\cite{shao2018spread} or other malicious actors~\cite{weedon2017information}, who often target elites, like celebrities, pundits, or politicians. From there, false claims rebroadcast by these elites enjoy further dissemination, reaching even wider audiences.
 
\begin{figure}[tbh]
  \vspace{1em}
  \includegraphics[width=\columnwidth]{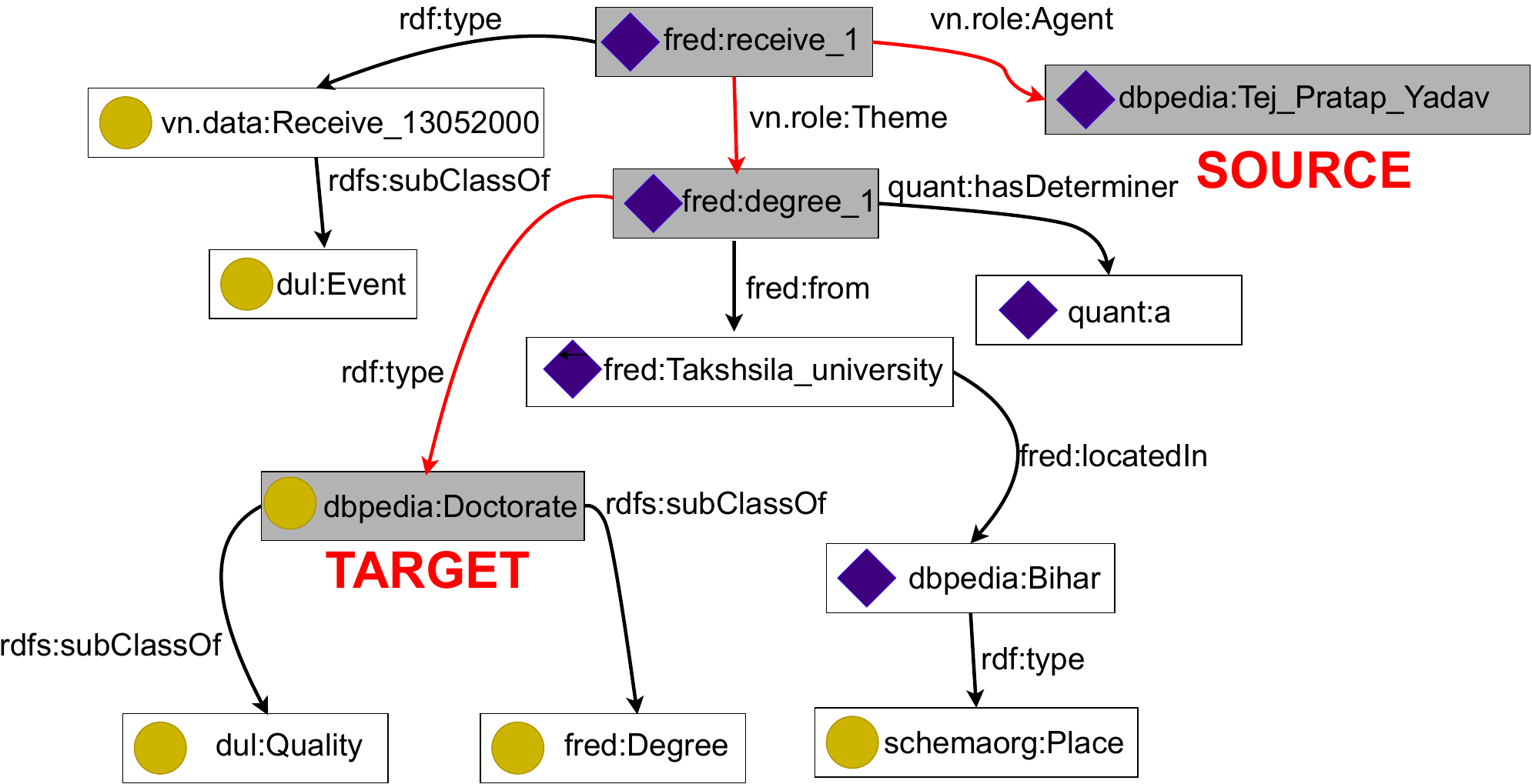}
  \caption{Schematic example of our approach. The RDF graphlet generated by a machine-reading tool (FRED) for the claim ``\emph{Tej Pratap Yadav receives a doctorate degree from Takshsila University in Bihar}'' (a known misinformation claim~\cite{kale2018no}). The shortest undirected path between the source (\texttt{dbpedia:Tej\_Pratap\_Yadav}) and target (\texttt{dbpedia:Doctorate}) is shown in red. The nodes along the path are highlighted in gray.}
  \Description{Schematic example of our approach. The RDF graphlet generated by a machine-reading tool (FRED) is visualized for the sentence ``\emph{Tej Pratap Yadav receives a doctorate degree from Takshsila University in Bihar}'' (a known misinformation claim~\cite{kale2018no}). There are a series of nodes representing different ontological entities (i.e. fred:Receive, schemaorg:Place), and directed edges between them, labeled with their ontological relation (i.e. \texttt{owl:sameAs}). The nodes along the shortest undirected path between \texttt{dbpedia: Doctorate} and \textt{dbpedia:Tej_Pratap_Yadav} are highlighted in gray.}
  \label{fig:fred}
\end{figure}

Misinformation has become an emerging focus of computational social scientists seeking to understand and combat it~\cite{ciampaglia2018fighting,Vlachos2014}. Network analysis and natural language processing (NLP) provide insight into the community organization and stylistic patterns that are indicative of misinformation, respectively, however they often fail to engage with the ideological content being shared. Online discourse typically takes the form of unorganized and unstructured data which is a significant limiting factor to performing content analysis. Existing work on semantic ontologies and knowledge base development has proved to be a guiding method in structuring online information. A knowledge base most commonly structures knowledge in the shape of semantic triples; a semantic triple is composed of two entities (e.g. a person, place, or thing) and a predicate relation between them. An example of a semantic triple is \texttt{<Washington\_D.C, capitalOf, United\_States\_of\_America>}. This structure allows for concepts to be reduced to machine-readable data which can be compiled into traversable (and understandable) networks of information. The result is a data structure that can be used to provide quantitative analysis of online discourse.

An example of knowledge bases application in combating misinformation regards computational fact-checking. Fact-checking is recognized as an antidote to misinformation~\cite{lewandowsky2012misinformation}, especially with respects to claims spread by political elites. For example, \citet{nyhan2015effect} show that alerting politicians to the risk of being fact-checked leads to less inaccuracy and better ratings. Unfortunately, fact-checking claims at the scale of the web is a hard task. A fact-checker must first identify claims that are worthy of being checked, then they must research the claim~\cite{silverman2014verification, borel2016chicago}, and finally write, publish, and circulate their conclusion on the web. In general, there is a lag of approximately 15 hours between the consumption of misinformation and the appearance of corrections~\cite{shao2016hoaxy}. The time investment required of human fact-checkers leads to an open opportunity for the development of many automated fact-checking~\cite{wu2014toward,ciampaglia2015computational} or verification~\cite{liu2015realtime} strategies. One approach is based on identifying missing relations in structured knowledge bases~\cite{lao2011random, ciampaglia2015computational,shiralkar2017finding,Shi2016}.  This approach takes a claim in the form of a semantic triple and checks its validity against the sets of triples in the knowledge base that connect the subject and object. When the knowledge base is viewed as a network, this task is equivalent to link prediction~\cite{liben2007link}. 

This approach has proven very promising, but its main restriction is that of its input. Modeling a claim using semantic triples is a nontrivial task, and has limited the application of such an approach. It requires choosing a semantic ontology (or developing a new one) which is able to model claims in a consistent and non-redundant manner. Once an ontology has been established, the next step is relation extraction --- the task of reducing a text into a semantic triple that both captures the meaning and fits within the ontology. This task is challenging when addressing a compound factual claim with many subjects and relations; this challenge is amplified when considering a claim that may contain sarcasm, opinion, humor, or any other nuance of language that can be present in online discourse. 

In this paper, we present a novel relation extraction method built upon semantic dependency trees, see Figure~\ref{fig:fred} for a schematic example. Our approach to the problem is based on the intuition that knowledge of the nodes and relations along the path between the subject and object of a triple (e.g.~\texttt{Washington,\_D.C.}, and \texttt{United\_States\_of\_America}) provides useful information that can be leveraged for extracting its relation (i.e.~\texttt{capitalOf}). This well-established phenomenon was first observed by~\citet{bradley1992learning}. Later, \citet{Bunescu2005} used it in the context of a kernel-based approach. Here, we take advantage of recent advances in graph representation learning to overcome the above challenges posed by online discourse in applying such an approach. Specifically, we parse a large corpus of Wikipedia snippets, annotated with information about one of 5 relations from the DBpedia ontology,
combine the resulting dependency trees into a larger semantic network, and finally use node embedding techniques to obtain a high-dimensional representation of this corpus-level network. We find that graph traversal in this learned representation provides a strong signal to discriminate between multiple possible relations. 

This approach allowed us to effectively extract these relations in natural language (extraction accuracy measured as the area under the ROC curve, $\mathrm{AUC}=0.976$). We then tested this model's ability to generalize to a set of real-world claims (reviewed by professional fact-checkers and annotated using the ClaimReview~\cite{guha2016schema.org} schema), obtaining again a very good signal (extraction $\mathrm{AUC}=0.958$). 

\begin{figure}
  \centering
  \includegraphics[width=\columnwidth]{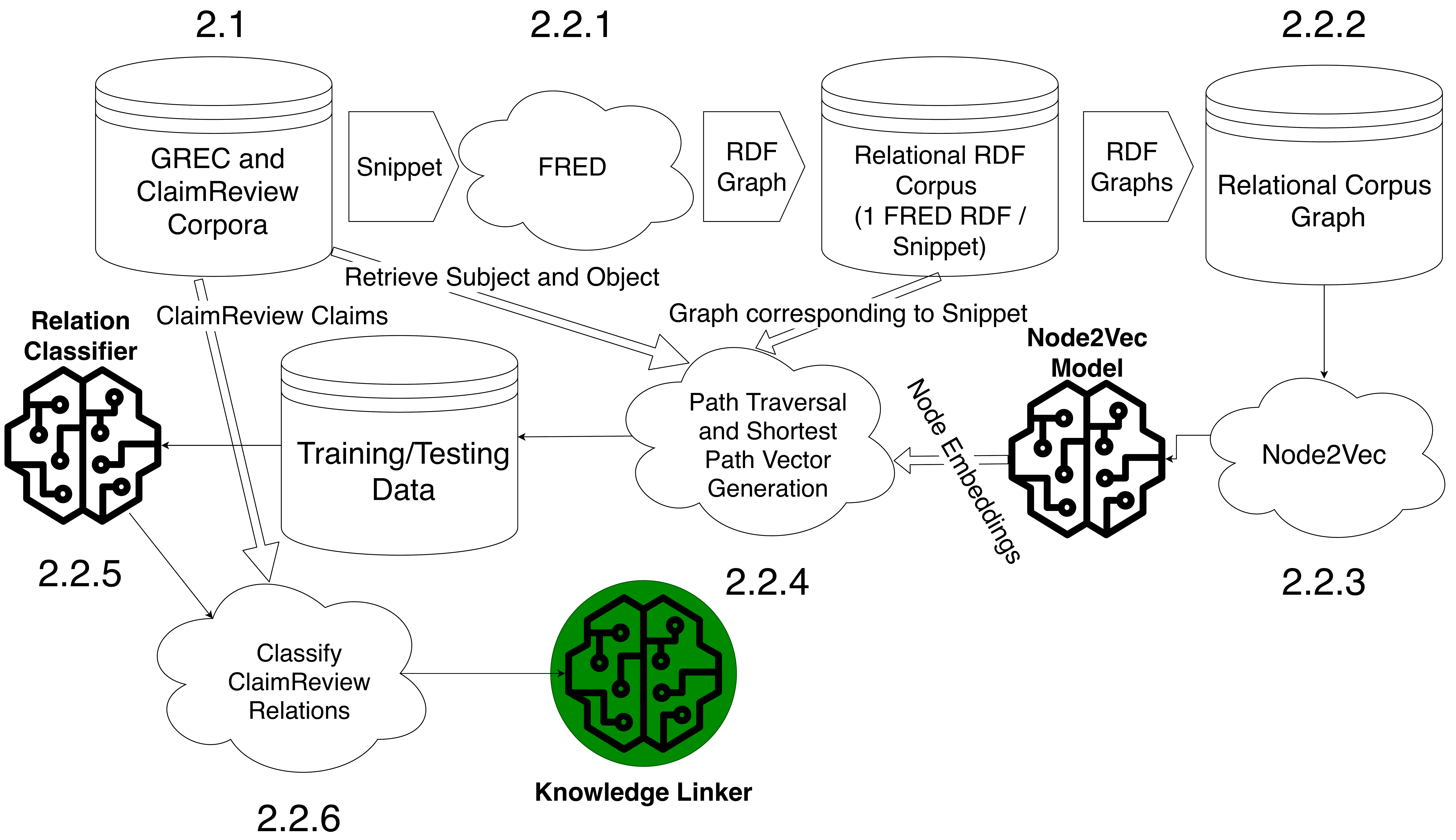}
  \caption{Schematic illustration of an integrated extraction and verification pipeline using our relation extraction tool REMOD. The white components correspond to the various steps needed to perform relation extraction. Numbered labels correspond to section headings in the manuscript. To show the potential for integration with external tools, as an additional step in the pipeline the green node shows the use of an off-the-shelf fact-checking algorithm~\cite{ciampaglia2015computational}.}
  \Description{This figure visualizes the steps in the proposed pipeline, with each step in the pipeline labeled with the corresponding section in which the description of that step occurs. Section 2.1 is the corpus, which gets passed to FRED (Section 2.2.1) which generates RDF graphs. From here, the RDF graphs get stitched into a full corpus graph (Section 2.2.2) and the Node2Vec algorithm is used to embed the nodes (Section 2.2.3). These embeddings are used to calculate vectors for the shortest paths between nodes in the RDF graphs (Section 2.2.4) and then used as input into a machine learning model for relation classification (Section 2.2.5). Finally, relevant claims are extracted from the ClaimReview corpus and passed into an off-the-shelf fact-checking algorithm like knowledge linker  (Section 2.2.6).}
  \label{fig:pipeline}
\end{figure}

As an example of a potential application of this technique, we show that, thanks to our method, a wider range of online discourse samples is amenable to analysis than before. In particular, we integrate our approach into a pipeline (see Figure~\ref{fig:pipeline}) that uses off-the-shelf fact-checking algorithms to analyze a subset of ClaimReview-annotated online discourse samples. Using this pipeline, we obtain very encouraging results on two separate tasks: First, on samples of `simple' online discourse claims, which can be effectively summarized (and thus fact-checked) by extracting a single RDF triple, we outperform a claim-matching baseline based on state-of-the-art representation learning (verification $\mathrm{AUC}=0.833$). Second, on more complex claims, 
from which one can extract multiple relevant relations, and therefore cannot be fact-checked directly, the fact-checker can still identify evidence in support or against the claim with good accuracy (verification $\mathrm{AUC}=0.773$). 

The rest of this paper is structured as follows: Section~\ref{sec:methods} details the datasets used, as well as the methods used in the various steps of the pipeline. Section~\ref{sec:results} shows the results of both the relation classification task and the fact checking tasks. Section~\ref{sec:relatedwork} goes into detail on relevant prior work from the literature on relation classification, misinformation detection, and computational fact-checking. Finally, Section~\ref{sec:discussion} discusses the impact and importance of our results, as well as addresses methods that may be used to improve upon this work in the future.

\section{Methods}\label{sec:methods}

Our relation extraction pipeline is described in Figure~\ref{fig:pipeline}. Roughly speaking, the main task of our pipeline is a supervised relation extraction task (white nodes), but since later we show how this task can be integrated to perform an additional unsupervised fact-checking, in the figure we show also this final step (green node). Collectively these two tasks leverage a number of different data sources, so we start by describing the various datasets used in building the pipeline. We then describe the various components of the pipeline proper.   

\subsection{Datasets}

For the main relation extraction task, we use two main corpora, both compiled by Google: the Google Relation Extraction Corpus (GREC) and the Google Fact Check Explorer corpus, described below.

\subsubsection{Google Relation Extraction Corpus (GREC)}
The dataset of relations used was the Google Relation Extraction Corpus (GREC)~\cite{orr_2013}. This dataset contains text snippets extracted from Wikipedia articles that represent a subject/object relation, which can be described by the following defining questions:
\begin{description}
    \item[Institution] ``What educational institution did the subject attend?''
    \item[Education] ``What academic degree did the subject receive?''
    \item[Date of Birth (DOB)] ``On what date was the subject born?''
    \item[Place of Birth (POB)] ``Where was the subject born?''
    \item[Place of Death (POD)] ``Where did the subject die?''
\end{description}

Each entry in the dataset consists of a natural language snippet of text, the URL of the Wikipedia entry from which the text was pulled, the Freebase predicate, a Freebase ID for subject and object, and the judgements of five human annotators on whether the snippet does or does not contain the relation (some annotators also voted to "skip", representing no decision either way).
Freebase has been replaced with the Google Knowledge Graph since this dataset was generated, which limited the use of this dataset in its original form. We made a set of addenda\footnote{\url{https://github.com/mjsumpter/google-relation-extraction-corpus-augmented}} to the GREC to update it to be more machine-ready for current relation extraction tasks and knowledge bases. The addenda include the following for each entry: text strings for both subject and object, DBpedia URI for both subject and object, Wikidata QID for both subject and object, a unique identifier, and the majority annotator vote.

\begin{table}
\caption{Number of snippets per relation before and after filtering the GREC corpus.}
\begin{tabular}{lrrr}
\toprule
{} &  Total &  Retained & \% Retained\\
\midrule
Institution & $42,628$ & $19,900$ & $46.7$ \\
Education & $1,850$ & $806$ & $43.6$ \\
Date of Birth & $2,490$ & $1,010$ & $40.6$ \\
Place of Birth & $9,566$ & $4,005$ & $41.9$ \\
Place of Death & $3,042$ & $1,307$ & $43.0$ \\
\bottomrule
\end{tabular}
\label{table:snip_len}
\end{table}

\begin{figure*}
  \includegraphics[width=\textwidth]{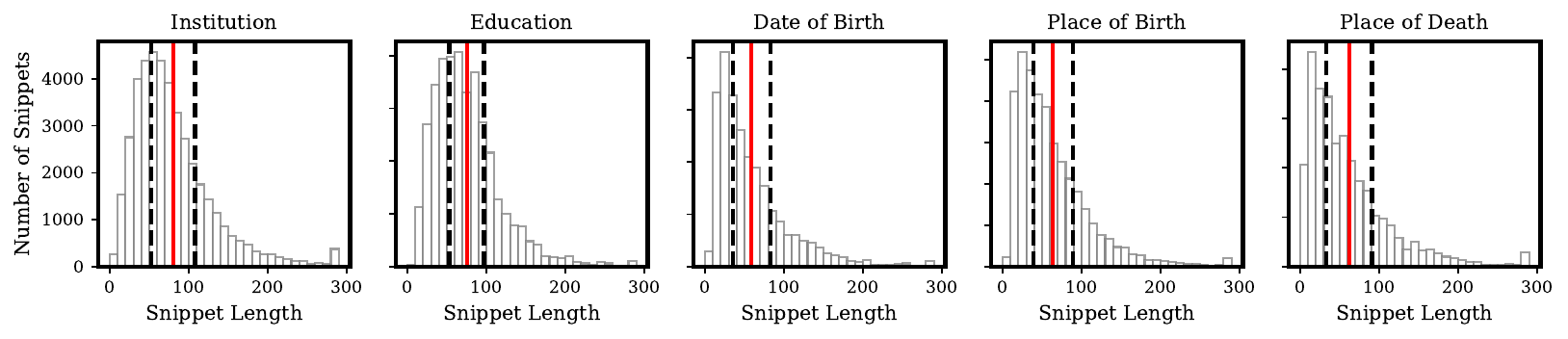}
  \caption{Distribution of snippet lengths found in the GREC. The red solid line corresponds to the average snippet length (in words) and the dashed lines to $\pm 0.5\sigma$ of the average. Snippets were kept if they were within this interval.}
  \Description{An image showing five histograms side-by-side. Each histogram represents the distribution of word lengths found in the snippets of each relation in the Google Relation Extraction Corpus. A solid red line is drawn at the mean of each distribution, and two dotted black lines demarcate 0.5 standard deviations from the mean.}
  \label{fig:snippet_hist}
\end{figure*}

The snippets varied considerably in length. The distribution of word lengths can be found in Figure~\ref{fig:snippet_hist}. Because we relied on a third-party API to parse the snippets, to reduce potential bias due to snippet length, and to ensure only the most characteristic relations were modeled, snippets were removed if they were not within ${\pm 0.5}$ standard deviations of the mean snippet length (measured in words), per relation. Table \ref{table:snip_len} shows the number of snippets retained, per relation.

\subsubsection{Google Fact Check Explorer Corpus}
Researchers at Duke University and Google have developed an annotation standard named ClaimReview~\cite{schema.org} to help annotate structured fact-checks on the web. It allows fact-checkers to add structured markup to their fact-checks with info that identifies distinct properties of a claim (i.e. claim reviewed, the rating decision, the source, etc.). This semi-structured data allows fact-checks to be catalogued and queried by search engines. The Google Fact Check Explorer tool\footnote{\url{https://toolbox.google.com/factcheck/explorer}} collects all the ClaimReview fragments published by fact-checking organizations that meet a set of established guidelines\footnote{\url{https://developers.google.com/search/docs/data-types/factcheck\#guidelines}}, which are the same standards for accountability, transparency, and accuracy used by Google News to select publishers.
We collected claims from the Google Fact Check Explorer tool up until 04/2020. From this corpus, we produced a dataset of 49,770 ClaimReview-annotated claims. 
Of the 20,817 English claims in the dataset, we searched for claims that contained one of the relations represented in the GREC, using WordNet~\cite{fellbaum1998wordnet} synonyms to select search terms (see Table~\ref{table:wordnet}). This procedure yielded a subset of 28 claims that met this criteria.

\begin{table}
\caption{The set of WordNet synonyms used to extract relevant claims from the ClaimReview database}
\begin{tabular}{p{2cm}p{5.5cm}}
\toprule
{} &  WordNet synonyms per relation\\
\midrule
Institution    &   \textit{attend, university, college, graduate} \\
Education      &   \textit{graduate, degree} \\
Date of Birth  &   \textit{born, born on} \\
Place of Birth &   \textit{born, birthplace, place of birth, place of origin} \\
Place of Death &   \textit{deceased, died, perished, passed away, expired }\\
\bottomrule
\end{tabular}
\label{table:wordnet}
\end{table}

\subsection{REMOD}
The main contribution of this work is REMOD (which stands for Relation Extraction for Modeling of Online Discourse), a novel tool for relation extraction that extract RDF triples from semi-structured samples of online discourse. To do so, our tool leverages an annotated corpus of past claims and relations. In the example pipeline shown in Figure~\ref{fig:pipeline}, the various steps of REMOD correspond to the white nodes, which we describe in more detail below. (The figure is labeled with numbers corresponding to the following section numbers, which elaborate on each step of the process.) 
To facilitate the replication of our results, the source code of REMOD is freely available online at \url{https://github.com/mjsumpter/remod}. 

\subsubsection{Semantic Parsing}
Our workflow begins with natural language snippets. To parse these snippets we used FRED, a machine reading tool based on Discourse Representation Theory and linguistic frames~\cite{Gangemi2017}, described by the authors as ``semantic middleware''. FRED is an NLP tool that combines frame detection, type induction, named-entity recognition, semantic parsing, and ontology alignment, all into a single tool. The authors provide a RESTful API to access it. When provided with a text string as input, it returns a Resource Description Framework (RDF) graphlet of the semantic parse tree of the input. (In practice, FRED produces DAGs instead of trees due to entity linking to external ontologies, hence our referring to them as `graphlets'.) An example of these RDF graphlets is shown in Figure \ref{fig:fred} for the ClaimReview snippet of a known misinformation claim~\cite{kale2018no}.

\subsubsection{Corpus Graph Composition}

\begin{figure}
  \centering
  \includegraphics[height=4.3cm]{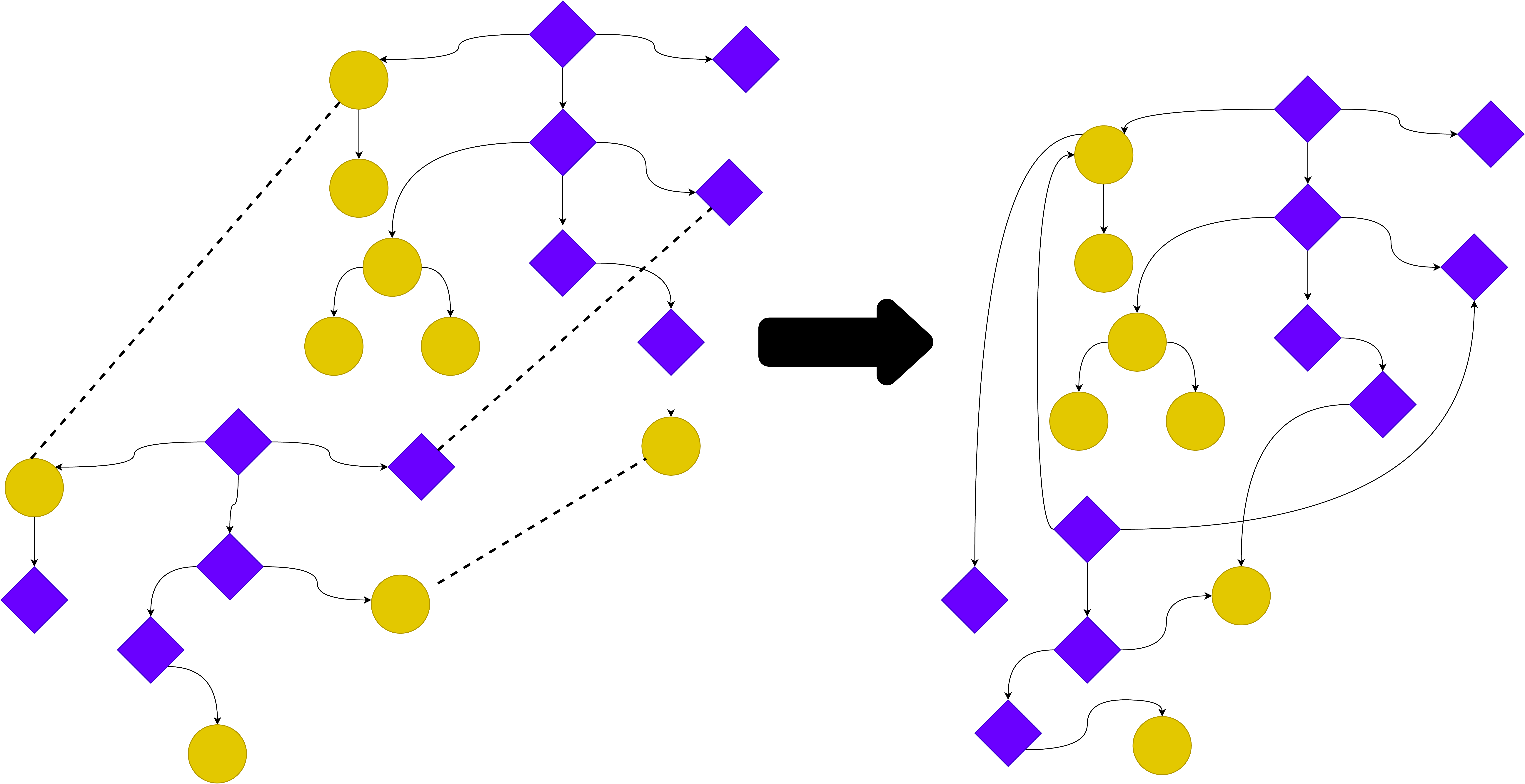}
  \caption{A visualization of how two separate RDF graphlets were stitched together along identical nodes.}
  \Description{This figure depicts two RDF graphlets on the left side, with identical nodes paired using dotted lines. A bold arrow then points to a single RDF graphlet on the right side, which depicts a single RDF graphlet that is the combination of the two RDF graphlets on the right side, but with their identical nodes combined.}
  \label{fig:stitch}
\end{figure}

In a realistic environment, many claims of different relations will exist in the same corpus. To mimic this environment, we composed a single `corpus' graph, which was composed of every FRED RDF graphlet generated from the corpus snippets. For named entities, FRED defaults to generating nodes for its own namespace (e.g.~\texttt{fred:Doctorate}), then if it finds that the same entity is present in an existing ontology, it links to that ontology (e.g.~\texttt{dbpedia:Doctorate}). Since these equivalent entities were redundant, we contracted the two nodes into a single vertex, and use the URI from the linked ontology (i.e.~DBpedia in this example) as its new URI. The corpus graph was than created by stitching together all the contracted RDF graphlets: if two graphlets share one or more nodes (i.e. two or more nodes have the same URI), then we consider the union of the two graphlets, and contract any pair of such nodes into a single node. This new node is incident to the union of all incident edges in the two original graphlets. An example of this is shown in Figure~\ref{fig:stitch}. The resulting corpus graph consists of $212,976$ nodes and $832,367$ edges.

\subsubsection{Node Embedding}\label{sec:params}

The corpus graph is effectively a combined semantic parse tree of the selected snippets from the corpus. To better exploit this structure in machine learning tasks, we generated node embeddings using the Node2Vec algorithm~\cite{Grover2016}. Node2Vec generates sets of random walks for each node, which are then substituted in place of natural language sentences as input into the Word2Vec model. There are two important parameters which will influence the nature of the embeddings: the return parameter $p$ and the in--out parameter $q$. For $p>1$ there is a higher likelihood of returning to a visited node in the random walks, whereas for $q>1$ there is an increased likelihood of exploring unvisited nodes.
We performed a grid search of $p$ and $q$ parameters (see Section~\ref{ssec:relation_class}), and determined the best choice for these parameters to be $p=2$ and $q=3$; this configuration captures what the authors of Node2Vec call the `global' topological structure of the graph.
The other parameters of Node2Vec were chosen as follows: the dimension of the vector space was set to $256$; the number of walks to $200$; the walk length to $200$; and, finally, the context window to $50$.

\subsubsection{Path Traversal for Finding Relations}
Our approach is inspired by the well-known idea that finding paths over structured knowledge representations can help learning new concepts~\cite{bradley1992learning}. More recently, Bunescu and Mooney~\cite{Bunescu2005} confirmed the intuitive conclusion that the shortest path between entities in a dependency tree captures the significant information contained between them. Therefore, we sought to develop a classifier that could distinguish between the shortest paths of different semantic relationships. To do so, for each snippet in the corpus, the subject and object were retrieved, along with the original (i.e., non-stitched) RDF graphlet of that specific snippet.
The nodes corresponding to the subject and object were identified in the RDF graphlet. With the terminal nodes identified, the shortest path in the original RDF graphlet was calculated (Figure~\ref{fig:fred}). Finally, we generated a final embedding by averaging along the path:
\[
\frac{1}{n} \sum_{i=1}^n \vec{v_i} 
\]
where $v_1, \ldots, v_n$ is a path and $\vec{v}\in \mathbb R^d$ is the vector associated to $v\in V$. This resulted in a final vector representing the aggregated sequence of nodes along the shortest path between subject and object. 

\begin{figure}
  \includegraphics[width=\columnwidth]{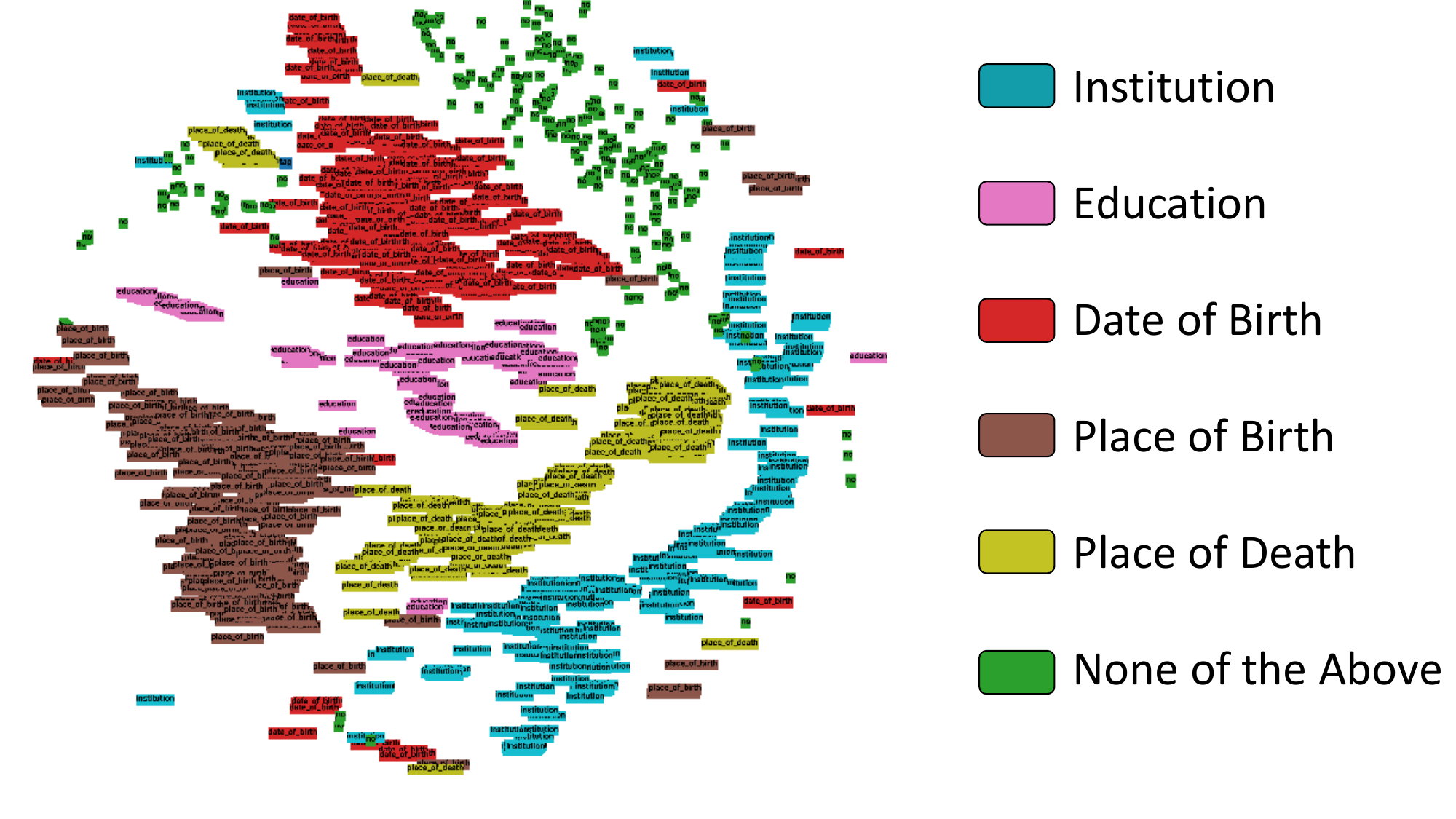}
  \caption{The shortest path vectors of GREC relations projected into 2D using t-SNE. Each color represents a different semantic relation, with a sixth color to mark snippets for which a majority of annotators voted `No (relation)'.}
  \Description{A visualization of many colored labels, each pertaining to a different vector representing the shortest path. Each color represents a different relation, and green represents snippets that had none of the selected relations. There is relatively clear clusters of similar colors in this image.}
  \label{fig:tsne}
\end{figure}

This process resulted in a $256$-dimensional vector for each snippet in the corpus. All results shown in the next section were obtained from these vectors. We projected the vectors into a lower-dimensional space using t-SNE. The visualization of these vectors is shown in Figure~\ref{fig:tsne}, where each color corresponds to a different relation. The projection reveals a good separation of vectors based on the relation they represent.

\subsubsection{Relation Classification}

We trained several classification models on the resulting set of shortest path vectors. The selected classifiers were Logistic Regression, $k$-NN, SVM, Random Forest, Decision Tree, and a Wide Neural Net.
Samples that were rated by the annotators to not contain a specified relation were removed, and then the dataset was balanced to the lowest frequency class (Education, $N=598$ samples). Readers will note this is a decrease from the $806$ reported in Table \ref{table:snip_len}; FRED was not always accurate at identifying entities and occasionally returned corrupted RDF graphs, resulting in a small loss of data. To effectively compare different classifiers, training was done using a 64\%/16\%/20\% training/validation/testing split. This resulted in a final training dataset of $1,913$ samples (5 classes, $N\approx{}382$ samples/class), with a validation set of $479$ samples, and an additional $598$ samples held for testing. The $28$ selected ClaimReview claims were held as an additional test set, which is elaborated on in Section \ref{ssec:fact_check}.


\subsubsection{Fact-Checking}\label{ssec:fact_check}

To demonstrate the usefulness of our method, we show that REMOD can be integrated as the first step of a fact-checking pipeline using existing, off-the-shelf tools to verify online discourse claims annotated using the ClaimReview standard. To perform fact-checking, we rely on the work of Shiralkar et al.~\cite{shiralkar2017finding}, who provide open-source implementations of several fact-checking algorithms\footnote{\url{https://github.com/shiralkarprashant/knowledgestream}}. These algorithms can be used to assess the truthfulness of a statement, but of course any tool that takes RDF triple in input could be used as well. 
To extract relations from ClaimReview snippets, we used the deep neural network classifier, which was the most successful classifier from the prior step, and feed the extracted triples into the fact-checker.

Of course, when integrating two distinct tools one has to make sure that any error originating in the first tool does not affect the performance of the second tool. Therefore, to avoid cascading errors we removed some claims from our dataset. We removed two types of errors. First, we removed any claim where the relation was misclassified, to avoid feeding inaccurate inputs into the fact-checker. Second, FRED is not always able to link both the subject and object entities to DBpedia, which is a requirement for using the fact-checking algorithms of~\citet{shiralkar2017finding}. Thus we also removed claims that did not have both the subject or object linked to the DBpedia ontology. Of the original 28 claims, this filtering resulted in 13 remaining ClaimReview claims used in our evaluation.

Additionally, we also manually checked whether the overall claim reduces to the extracted triple (in the sense that verifying the triple also verifies the overall claim). This distinction is important since it allows us to gauge the ability of our system to check entire claims automatically, in a purely end-to-end fashion. Finally, these remaining claims were passed as input to three fact-checking algorithms: Knowledge Stream, Knowledge Linker, and Relational Knowledge Linker~\cite{shiralkar2017finding}. 

As a baseline, we trained a Doc2Vec model \cite{le2014distributed} on the entirety of the ClaimReview corpus, and used this model to fact-check statements by matching them with other similar claims. In particular, given an input claim, to produce a truth score with the baseline model we ranked all claims in the ClaimReview corpus by their similarity and averaged the truth scores of the top $k$ most similar matching claims. 
We removed fact-checking organizations that used scaleless fact-check verdicts (i.e. factcheck.org); for those that had scales, we assigned truth scores to every claim, setting "False" to a baseline of 0, unless a scale explicitly stated a different baseline (i.e. PolitiFact ranks "Pants on Fire" lower than "False").

\section{Results}\label{sec:results}

\subsection{Graph Representation}

The corpus graph is composed of dependency trees, and so the corpus graph is naturally a directed graph; edges are also all weighted equally. This design has a strong influence on path traversal, since directed edges reduce the number of available paths and the cost of taking an edge (or its absence) influences the choice of one path over another. For completeness, we considered all four combinations of taking either a directed or undirected graph, and of having edge weights or not. 
Let $v_i,v_j\in V$ represent two nodes in the dependency graph that are incident on the same edge. The weight $w_{ij}$ between them is the angular distance between the respective node embeddings: 
\[
w_{ij} = \frac{1}{\pi}\arccos\left(\frac{\vec{v}_i \cdot \vec{v}_j}{\|\vec{v}_i \|\cdot \|\vec{v}_j \|}\right)
\]
Where $\vec{v}$ is the vector associated to $v\in V$.

Table~\ref{table:graph_style} shows that the undirected, unweighted graph yields the best classification results, which prompts two observations. The first is that directed edges reduce the number of available pathways to connect two nodes. Second, and perhaps a bit surprisingly, we observe that the unweighted network performs better than the weighted one. Because node embeddings were the same in the two variants, the final feature vector used for relation classification would be different only if a different shortest path was found. This could be possible if edges that are more relevant to discriminating the relation were assigned large weights, compared to other, less relevant edges. 

\begin{table}
\caption{AUC of Wide DNN on the relation classification task using different types of graph to represent the corpus graph.}
\begin{tabular}{rrr}
\toprule
\textbf{AUC} & \textbf{Unweighted} & \textbf{Weighted} \\
\midrule
Undirected & \textbf{0.976} & 0.964 \\
Directed & 0.966 & 0.967 \\
\bottomrule
\end{tabular}
\label{table:graph_style}
\end{table}

\subsection{Classification for Relation Extraction}\label{ssec:relation_class}

The results of the relation classification task are shown in Table \ref{table:diff_models}. The outcome of these various tests reveal that the node embeddings do contain information regarding the semantic nature of the GREC relations, however they are not neatly separable by decision planes. It is notable that models we tested are often more successful in precision than in recall. This suggests that the more complex model, such as a deep neural network (DNN), is necessary to identify the less characteristic samples of a relation. 
To improve these results, we performed a grid search on the Node2Vec $p$ and $q$ parameters (with values of $0.25$, $0.5$, $1$, $2$, $3$, and $4$). The best overall results were a product of a `global' configuration, using $p = 2$ and $q = 3$, which achieved an AUC of $0.976$ on the test set. 
%
%
To evaluate our method, as a baseline we generated 300-dimension vectors for each snippet from a Word2Vec model, pre-trained on Wikipedia~\cite{yamada2016joint}. This is the same source of the GREC corpus, which provided training data for model. These embeddings were then used as features to train a DNN and a logistic regression model for relation extraction. REMOD shows a marked improvement in both instances, indicating an effective approach to relation extraction.
%

\begin{table}
\caption{Results of the relation classification task using different ML models, on an unweighted, undirected corpus graph, as compared to training with Word2Vec embeddings.}
\begin{tabular}{lrrrr}
\toprule
{} &  Precision &  Recall & F1 & AUC \\
\midrule
Decision Tree & 0.64 & 0.64 & 0.64 & 0.773 \\
Random Forest & 0.81 & 0.67 & 0.61 & 0.793 \\
$k$-NN        & 0.78 & 0.74 & 0.74 & 0.841 \\
SVM           & 0.81 & 0.77 & 0.77 & 0.855 \\
Log. Regr.    & 0.80 & 0.71 & 0.71 & 0.827 \\
Wide DNN      & \textbf{0.85} & \textbf{0.85} & \textbf{0.85} & \textbf{0.976} \\
\midrule
Word2Vec+Log. Regr.& 0.66 & 0.47 & 0.44 & 0.658 \\
Word2Vec+Wide DNN  & 0.61 & 0.63 & 0.61 & 0.883 \\
\bottomrule
\end{tabular}
\label{table:diff_models}
\end{table}

\subsection{Extraction of ClaimReview Relations}

Table~\ref{table:claimreview} in the appendix shows the claims selected from the ClaimReview corpus, in addition to the relation they contain ("Actual"), the relation predicted by REMOD ("Predicted"), the truth rating as determined by a fact-checker ("Rating"), and whether verifying the relation is equivalent to verifying the claim ("Claim $\equiv$ Triple"). The AUC of the predicted relations is $0.958$. Inspecting the misclassified samples, we see that REMOD made mistakes between similar relations (e.g. place of birth and date of birth), which often occur in similar sentences.

\subsection{Fact-Checking}

We next test the integration with fact-checking algorithms. In particular, we use the fact-checker for two similar, but conceptually distinct tasks: 1) fact-checking an entire claim (\emph{fact-checking}), and 2) identifying evidence in support or against a claim (\emph{fact verification}).
For example, for claim~\#1 (see Table~\ref{table:claimreview}), Penny Wong was indeed born in Malaysia, even though the assertion that she is ineligible for being elected into the Australian parliament is false. Thus, in this case the extracted triple is only additional evidence, but is not able in itself to capture the entire claim. 
We manually fact-checked all the extracted relations, and compared their truth rating with the one provided by the human fact-checker for the whole claim. Table~\ref{table:claimreview} lists this information under the column ``Claim $\equiv$ Triple'', which is true (indicated by a checkmark) when the extracted relation summarizes the whole claim (e.g. claim~\#3). This distinction is important: as mentioned before, although our relation extraction pipeline is capable of predicting a relation for all the entries in Table~\ref{table:claimreview}, not all triples that are correctly predicted can be fed to the fact-checking algorithms, due to incomplete entity linking. For the task of identifying supporting evidence, we find a total of 13 ClaimReview claims that are amenable to fact-checking. For the task of checking an entire claim, this number is further reduced to 7 claims.   

\subsubsection{Fact Verification}

Table~\ref{table:relation_fact_check} shows the results of verifying individual pieces of evidence in support or against any of the 13 ClaimReview claims identified by REMOD, using any of the three algorithms for fact-checking RDF triples. Relational Knowledge Linker and Knowledge Stream were the best performers. Note that since our baseline is intended to emulate a true fact-checking task, in this case we do not run the baseline since the similarity is based on the whole claim, and thus would not be a meaningful comparison with our method, which focuses only on a specific relation within a larger claim. 

\begin{table}
\caption{The performance of the fact-checking algorithms on predicting the validity of the relations.}
\begin{tabular}{lr}
\toprule
\textbf{Method} & \textbf{AUC} \\
\midrule
Knowledge Linker & 0.636 \\
Relational Knowledge Linker & \textbf{0.773}\\
Knowledge Stream & \textbf{0.773}\\
\bottomrule
\end{tabular}
\label{table:relation_fact_check}
\end{table}

\subsubsection{Fact-Checking}

We test here the subset of claims for which checking the triple is equivalent to checking the entire claim. In this case, REMOD yields 7 claims that can be used as inputs to the fact-checking algorithms. Table~\ref{table:fact_check} shows the results of our 7 ClaimReview claims, on the three fact-checking algorithms, along with the baseline. Here, the baseline emulates fact-checking by claim matching. 

Since we are using claim-matching to perform fact-checking, we consider three different scenarios to make the task more realistic. In particular, we match the claim against three different corpora by higher degree of realism: 1) the full ClaimReview corpus (`All'), 2) all ClaimReview entries by PolitiFact only (`PolitiFact'), and 3) all ClaimReview entries from the same fact-checker of the claim of interest (`Same'). 
The first case (`All') is meant to give an upper bound on the performance of claim matching but is not realistic, since it makes use of knowledge of the truth score of potentially future claims, as well as of ratings for the same claim but by different fact-checkers. The second case (`PolitiFact') partially addresses this second unrealistic assumption by using only claims from a single source. Thus, it does not have access to truth scores by different organizations for the same claim, but it does still have access to future information. Both 1) and 2) can be thus regarded as gold standard measures of performance. The last one (`Same') is the more realistic one, since it emulates the scenario of a fact-checker who may check a claim for the first time, and who thus cannot have access to claims fact-checked afterwards nor by ratings of the same claim by different fact-checkers. In all three cases, the claim being matched was removed from the corpus, to prevent trivially perfect predictions. 
Relational Knowledge Linker and Knowledge Stream are still the best performing of the fact-checking algorithms and manages to reach, if not exceed, the performance of the gold standard (Claim Matching--All, or --PolitiFact).   

\begin{table}
\caption{Results of the fact-checking algorithms. (CM = Claim Matching; KL = Knowledge Linker; Rel.~KL = Relational Knowledge Linker; KS = Knowledge Stream.)}
\begin{tabular}{lrrrr}
\toprule
& $k=1$ & $k=3$ & $k=5$ & $k=10$ \\
\midrule
CM (All) & 0.417 & \textbf{0.625} & 0.500 & \textbf{0.625} \\
CM (PolitiFact) & 0.666 & 0.625 & \textbf{0.833} & 0.750 \\
CM (Same) & 0.500 & \textbf{0.583} & 0.25 & 0.25 \\
\midrule
KL & \multicolumn{4}{c}{0.500} \\
Rel.~KL & \multicolumn{4}{c}{\textbf{0.833}}\\
KS & \multicolumn{4}{c}{\textbf{0.833}}\\
\bottomrule
\end{tabular}
\label{table:fact_check}
\end{table}

\section{Related Work}\label{sec:relatedwork}

\subsection{Relation Extraction and Classification}

Relation extraction and classification is the task of extracting semantic relationships between two entities in natural language text and matching them to semantically equivalent or similar relations. This task is at the core of information extraction and knowledge base construction, as it effectively reduces statements to their core meaning; this is typically modeled as a semantic triple, (\textit{s,p,o}), where two entities (\textit{s} and \textit{o}) are connected with a predicate, \textit{p}. There are several distinct nuances and open challenges to effective relation extraction. Identifying attributes that discriminate between two objects provides a descriptive explanation to supplement word embeddings (i.e. lime is separated from lemon by the attribute `green'), and is currently most successful with SVM classifiers \cite{lai-etal-2018-sunnynlp}. Multi-way classification attempts to distinguish the direction of one-way relations (the \texttt{sonOf} relation is not bidirectional between two people), and has seen similar levels of success from solutions built with language models \cite{baldini-soares-etal-2019-matching}, convolutional neural networks \cite{wang-etal-2016-relation}, and recurrent neural networks \cite{xiao-liu-2016-semantic}. Distantly supervised relation extraction is a two-way approach whereby semantic triples are generated from natural language by aligning them with information already present in knowledge graphs \cite{xu-barbosa-2019-connecting}. Relation extraction performance is often assessed on the TACRED dataset~\cite{zhang2017tacred}. This is a large-scale dataset of $106,264$ examples used in the annual TAC Knowledge Base Population challenges, and covers $41$ relation types. The most successful solution to date is from~\citet{baldini-soares-etal-2019-matching}, who achieved a micro-averaged F1 score of $71.5\%$. Despite increasing availability of state-of-the-art machine learning architectures, relation extraction continues to be an open problem with much room for improvement.

\subsection{Knowledge Base Augmentation}

Knowledge base augmentation is a task that aims to add new relations to existing knowledge bases in an automated fashion \cite{weikum_info}. This task takes one of two approaches; the first infers new relations from existing triples in a knowledge base \cite{kb_enrich, socher_nn_kbaugment} --- this is essentially a link-prediction task that builds upon patterns found between entities in knowledge bases. The second approach mines data found on the web for knowledge discovery \cite{knowmore_yu, knowvault_Dong2014}. This approach relies on redundant relations found among the selected source materials, which may be as restrictive as Wikipedia articles \cite{paulheim_wikiaugment} or as extensive as the entire web \cite{knowvault_Dong2014}. Due to the potential for error based on the sources, \citet{dong_infotrust} developed a Knowledge-Based Trust (KBT) score for measuring the trustworthiness of selected sources. \citet{knowmore_yu} expand upon this by combining KBT scores with other entity/relation-based features to assign a unique score to each individual triple. 

\subsection{Detecting Information Disorder}

Information disorder is a catch-all term for the many kinds of unreliable information that one may encounter online or in the real-world~\cite{wardle2017information}, which includes disinformation, misinformation, fake news, rumor, spam, etc. Information disorder can also take on several modalities, including text, video, and images. The many varieties of information disorder make it challenging to develop any one approach for detection. This leads to a multi-model approach to detection based on three main modalities: the content of the information, the users who shared it, and the patterns of information dissemination on a network. Often bad content is generated by bots; this suggests that features captured from user profiles can be useful for distinguishing bots from humans~\cite{Shu2018}. Content detection is dependent on the medium; lexical features, sentiment, and readability metrics are used for text, while neural visual features are extracted from other content \cite{rashkin-etal-2017-truth, rubin-vashchilko-2012-identification, doi:10.1002/pra2.2015.145052010083}. Network detection methods model social media networks as propagation networks, measuring the flow of information \cite{shu2020hierarchical}. There has also been promising work into crowd-sourcing the task by allowing users to flag questionable content \cite{10.1145/3184558.3188722}. This task, while likely to remain imperfect, provides the important supplement of human supervision to all of the aforementioned tasks.

\subsection{ClaimBuster}

Hassan et al.~\cite{Hassan2017} released the first-ever end-to-end fact-checking system in 2017, called ClaimBuster. ClaimBuster is composed of several distinct components that work in sequence to accomplish the task of automated fact-checking. The first, \emph{claim monitor}, continuously monitors text published as broadcast television closed-captions, Twitter accounts, and as content on a selected set of websites. This text is passed to the \textit{claim spotter}, which scores every sentence by its likelihood to contain a claim that is worthy of fact-checking --- subjective and opinionated sentences receive a low score in this task. Once it has identified a set of check-worthy sentences, it uses a \emph{claim matcher} to search through fact-check repositories to return existing fact-checks that match the selected sentences. \emph{Claim checker} generates questions from the selected sentences and uses those questions to query Wolfram Alpha and Google to fetch supporting or debunking evidence as a supplement to the findings of \emph{claim matcher}. Finally, the \emph{fact-check reporter} builds a report from all of the gathered evidence that summarizes the findings of the ClaimBuster pipeline, and disseminates these findings through social media.

\subsection{Claim Verification}

Claim verification is arguably the key task of fact-checking --- to check a claim against existing evidence. It is related to the matching and checking subtasks of ClaimBuster, in that it is the task of checking whether a natural language sentence selected as evidence supports or debunks the correlated claim. To build out computational solutions to this task, datasets containing claims and their corresponding evidence are needed. There have been some datasets \cite{Vlachos2014, Fridkin2015, angeli2014naturalli} relevant to this task, however they are either not machine-readable or lacking in size. Thorne et al. \cite{Thorne2018} recognized this gap, and has since released a large-scale dataset to address these concerns, called FEVER. This dataset contains 185,445 claims with corresponding evidence that were manually classified as \texttt{SUPPORTED}, \texttt{REFUTED}, or \texttt{NOTENOUGHINFO}. This has been followed up with annual workshops that encourage participants to improve upon both the dataset and the claim verification task. The CLEF CheckThat! \cite{barroncedeno2020checkthat} series of workshops and conferences also seek to bring researchers together to improve claim verification, along with identifying and extracting checkworthy claims.

\subsection{Other Fact-Checking Methods}

Besides claim-matching approaches, there are a handful of existing algorithms for fact-checking, mostly based on exploiting content or characteristics of existing knowledge bases. Embedding approaches, such as TransE~\cite{Bordes2013}, seek to generate vector embeddings of knowledge bases, a task which is conceptually related to our approach. By generating these embeddings, they can perform link-prediction based on structural patterns of (\textit{s, p, o}) triples. In terms of a knowledge base, this amounts to adding new facts without any needed source material. For fact-checking, this approach can be used to test whether a triple extracted from a claim is a predicted link in the knowledge base; the pitfall of these methods, as with all embedding techniques, is they lack both interpretability and scalability. Other algorithms similarly consider paths within knowledge bases, but seek to address the interpretability problem. PRA~\cite{Lao2010}, SFE~\cite{gardner2015efficient}, PredPath~\cite{Shi2016}, and AMIE~\cite{Galarraga2013} all take the approach of mining possible pathways between two entities within a knowledge base. From these mined pathways, they generate sets of features to be used in supervised learning models for link-prediction. These have shown promise in their success at predicting the validity of a claim, however this also suffers from scalability. Knowledge bases that contain enough relevant information to be useful are very large, and path mining and feature generation becomes necessarily time-consuming. There are a few rule-based~\cite{ortona2018robust} methods for fact-checking, which rely on logical constraints of a knowledge graph and are naturally explainable. General, large-scale knowledge graphs do not have these logical constraints from which to build rules from, leaving this approach to fact-checking an open problem~\cite{Huynh2019}.

\subsection{Threats to Validity}

No method is perfect and our approach suffers from a number of limitations, which we briefly describe here. The main limitation of our pipeline lies in its discrete structure, which is prone to cascading failures.
Our main NLP tool, FRED, is a powerhouse of a tool and performed many important NLP tasks at once; however, it was not always completely accurate 
and many of our samples were returned as corrupted RDF graphs. Additionally, it was not always able to link the nodes to DBpedia, which limited the number of triples we could feed into our fact-checking algorithms. Cascading failures are common to many machine reading pipelines~\cite{mitchell2018never}. One way to overcome this issue would be to rely on a joint inference approaches~\cite{singh2013joint}. 
Another limitation of our methodology has to do with our use of distributed representations. For the task of fact-checking, the corpus is always growing; Node2Vec cannot generalize to unseen data and requires retraining. An inductive learning framework, such as GraphSAGE \cite{hamilton-graphsage}, can generate embeddings for unseen nodes, and is therefore a more practical algorithm for extending this pipeline. For the classification task, our machine learning models were relatively simple, and optimizing both the parameters and architecture of the neural network would likely see an increase in the accuracy and effectiveness of this method. Finally, a full evaluation of our method against transformer language models for relevant relation classification tasks~\cite{wu2019enriching} is left as future work.

\section{Discussion}\label{sec:discussion}

In this paper, we have presented a novel relation extraction algorithm and previewed its application when used to classify relations present in online discourse and automatically fact-check them against the information present in a general knowledge graph.
We developed a pipeline to facilitate the linkage of these two tasks. Our relation classification method leverages graph representation learning on the shortest paths between entities in semantic dependency trees; it was shown to be comparable to state-of-the-art methods based on a corpus of labeled relations ($\mathrm{AUC}=97.6\%$). This classifier was then used to reduce claims from online discourse to semantic triples with an AUC of $95.8\%$; these were used as input to fact-checking algorithms to predict the accuracy of the claim. We achieved an AUC of $83\%$ on our selected claims, which is at the least comparable to claim matching, but without the need for the corpus of existing claims that claim matching relies on. 

Our relation extraction method is a promising approach to distinguishing relations present in large online discourse corpora; scaling up this algorithm could provide an outlet for modeling online discourse within an established ontology. Additionally, our pipeline may serve as a proof-of-concept for future research into automated fact-checking. While it is a challenge to model all possible relations in a generalistic ontology like DBpedia, this pipeline could form the basis of tools for reducing the time needed to research an online discourse claim. 

\subsection*{Acknowledgements}


The authors would like to thank Google for making publicly available both the GREC dataset and the Fact Check Explorer tool, and Alexios Mantzarlis for feedback on the manuscript.

\bibliographystyle{ACM-Reference-Format}
\bibliography{works_cited}


\begin{thebibliography}{69}


\ifx \showCODEN    \undefined \def \showCODEN     #1{\unskip}     \fi
\ifx \showDOI      \undefined \def \showDOI       #1{#1}\fi
\ifx \showISBNx    \undefined \def \showISBNx     #1{\unskip}     \fi
\ifx \showISBNxiii \undefined \def \showISBNxiii  #1{\unskip}     \fi
\ifx \showISSN     \undefined \def \showISSN      #1{\unskip}     \fi
\ifx \showLCCN     \undefined \def \showLCCN      #1{\unskip}     \fi
\ifx \shownote     \undefined \def \shownote      #1{#1}          \fi
\ifx \showarticletitle \undefined \def \showarticletitle #1{#1}   \fi
\ifx \showURL      \undefined \def \showURL       {\relax}        \fi
\providecommand\bibfield[2]{#2}
\providecommand\bibinfo[2]{#2}
\providecommand\natexlab[1]{#1}
\providecommand\showeprint[2][]{arXiv:#2}

\bibitem[\protect\citeauthoryear{Allcott and Gentzkow}{Allcott and
  Gentzkow}{2017}]%
        {allcott2017social}
\bibfield{author}{\bibinfo{person}{Hunt Allcott} {and} \bibinfo{person}{Matthew
  Gentzkow}.} \bibinfo{year}{2017}\natexlab{}.
\newblock \showarticletitle{Social media and fake news in the 2016 election}.
\newblock \bibinfo{journal}{\emph{Journal of economic perspectives}}
  \bibinfo{volume}{31}, \bibinfo{number}{2} (\bibinfo{year}{2017}),
  \bibinfo{pages}{211--36}.
\newblock


\bibitem[\protect\citeauthoryear{Angeli and Manning}{Angeli and
  Manning}{2014}]%
        {angeli2014naturalli}
\bibfield{author}{\bibinfo{person}{Gabor Angeli} {and}
  \bibinfo{person}{Christopher~D. Manning}.} \bibinfo{year}{2014}\natexlab{}.
\newblock \showarticletitle{{N}atural{LI}: Natural Logic Inference for Common
  Sense Reasoning}. In \bibinfo{booktitle}{\emph{Proceedings of the 2014
  Conference on Empirical Methods in Natural Language Processing ({EMNLP})}}.
  \bibinfo{publisher}{Association for Computational Linguistics},
  \bibinfo{address}{Doha, Qatar}, \bibinfo{pages}{534--545}.
\newblock
\urldef\tempurl%
\url{https://doi.org/10.3115/v1/D14-1059}
\showDOI{\tempurl}


\bibitem[\protect\citeauthoryear{Baldini~Soares, FitzGerald, Ling, and
  Kwiatkowski}{Baldini~Soares et~al\mbox{.}}{2019}]%
        {baldini-soares-etal-2019-matching}
\bibfield{author}{\bibinfo{person}{Livio Baldini~Soares},
  \bibinfo{person}{Nicholas FitzGerald}, \bibinfo{person}{Jeffrey Ling}, {and}
  \bibinfo{person}{Tom Kwiatkowski}.} \bibinfo{year}{2019}\natexlab{}.
\newblock \showarticletitle{Matching the Blanks: Distributional Similarity for
  Relation Learning}. In \bibinfo{booktitle}{\emph{Proceedings of the 57th
  Annual Meeting of the Association for Computational Linguistics}}.
  \bibinfo{publisher}{Association for Computational Linguistics},
  \bibinfo{address}{Florence, Italy}, \bibinfo{pages}{2895--2905}.
\newblock
\urldef\tempurl%
\url{https://doi.org/10.18653/v1/P19-1279}
\showDOI{\tempurl}


\bibitem[\protect\citeauthoryear{Barron-Cedeno, Elsayed, Nakov, Martino,
  Hasanain, Suwaileh, and Haouari}{Barron-Cedeno et~al\mbox{.}}{2020}]%
        {barroncedeno2020checkthat}
\bibfield{author}{\bibinfo{person}{Alberto Barron-Cedeno},
  \bibinfo{person}{Tamer Elsayed}, \bibinfo{person}{Preslav Nakov},
  \bibinfo{person}{Giovanni Da~San Martino}, \bibinfo{person}{Maram Hasanain},
  \bibinfo{person}{Reem Suwaileh}, {and} \bibinfo{person}{Fatima Haouari}.}
  \bibinfo{year}{2020}\natexlab{}.
\newblock \bibinfo{title}{CheckThat! at CLEF 2020: Enabling the Automatic
  Identification and Verification of Claims in Social Media}.
\newblock
\newblock
\showeprint[arxiv]{2001.08546}~[cs.CL]


\bibitem[\protect\citeauthoryear{Bordes, Usunier, Garcia-Duran, Weston, and
  Yakhnenko}{Bordes et~al\mbox{.}}{2013}]%
        {Bordes2013}
\bibfield{author}{\bibinfo{person}{Antoine Bordes}, \bibinfo{person}{Nicolas
  Usunier}, \bibinfo{person}{Alberto Garcia-Duran}, \bibinfo{person}{Jason
  Weston}, {and} \bibinfo{person}{Oksana Yakhnenko}.}
  \bibinfo{year}{2013}\natexlab{}.
\newblock \showarticletitle{Translating Embeddings for Modeling
  Multi-relational Data}.
\newblock In \bibinfo{booktitle}{\emph{Advances in Neural Information
  Processing Systems 26}}, \bibfield{editor}{\bibinfo{person}{C.~J.~C. Burges},
  \bibinfo{person}{L.~Bottou}, \bibinfo{person}{M.~Welling},
  \bibinfo{person}{Z.~Ghahramani}, {and} \bibinfo{person}{K.~Q. Weinberger}}
  (Eds.). \bibinfo{publisher}{Curran Associates, Inc.}, \bibinfo{address}{Red
  Hook, NY, United States}, \bibinfo{pages}{2787--2795}.
\newblock


\bibitem[\protect\citeauthoryear{Borel}{Borel}{2016}]%
        {borel2016chicago}
\bibfield{author}{\bibinfo{person}{B. Borel}.} \bibinfo{year}{2016}\natexlab{}.
\newblock \bibinfo{booktitle}{\emph{The Chicago Guide to Fact-Checking}}.
\newblock \bibinfo{publisher}{University of Chicago Press},
  \bibinfo{address}{Chicago, IL, USA}.
\newblock
\showISBNx{9780226290935}
\showLCCN{2016018060}


\bibitem[\protect\citeauthoryear{Bovet and Makse}{Bovet and Makse}{2019}]%
        {bovet2019influence}
\bibfield{author}{\bibinfo{person}{Alexandre Bovet} {and}
  \bibinfo{person}{Hernán~A. Makse}.} \bibinfo{year}{2019}\natexlab{}.
\newblock \showarticletitle{Influence of fake news in Twitter during the 2016
  US presidential election}.
\newblock \bibinfo{journal}{\emph{Nature Communications}} \bibinfo{volume}{10},
  \bibinfo{number}{1} (\bibinfo{date}{Jan.} \bibinfo{year}{2019}),
  \bibinfo{pages}{7}.
\newblock
\showISSN{2041-1723}
\urldef\tempurl%
\url{https://doi.org/10.1038/s41467-018-07761-2}
\showDOI{\tempurl}


\bibitem[\protect\citeauthoryear{B{\"u}hmann and Lehmann}{B{\"u}hmann and
  Lehmann}{2013}]%
        {kb_enrich}
\bibfield{author}{\bibinfo{person}{Lorenz B{\"u}hmann} {and}
  \bibinfo{person}{Jens Lehmann}.} \bibinfo{year}{2013}\natexlab{}.
\newblock \showarticletitle{Pattern Based Knowledge Base Enrichment}. In
  \bibinfo{booktitle}{\emph{The Semantic Web -- ISWC 2013}},
  \bibfield{editor}{\bibinfo{person}{Harith Alani}, \bibinfo{person}{Lalana
  Kagal}, \bibinfo{person}{Achille Fokoue}, \bibinfo{person}{Paul Groth},
  \bibinfo{person}{Chris Biemann}, \bibinfo{person}{Josiane~Xavier Parreira},
  \bibinfo{person}{Lora Aroyo}, \bibinfo{person}{Natasha Noy},
  \bibinfo{person}{Chris Welty}, {and} \bibinfo{person}{Krzysztof Janowicz}}
  (Eds.). \bibinfo{publisher}{Springer Berlin Heidelberg},
  \bibinfo{address}{Berlin, Heidelberg}, \bibinfo{pages}{33--48}.
\newblock
\showISBNx{978-3-642-41335-3}


\bibitem[\protect\citeauthoryear{Bunescu and Mooney}{Bunescu and
  Mooney}{2005}]%
        {Bunescu2005}
\bibfield{author}{\bibinfo{person}{Razvan~C. Bunescu} {and}
  \bibinfo{person}{Raymond~J. Mooney}.} \bibinfo{year}{2005}\natexlab{}.
\newblock \showarticletitle{A Shortest Path Dependency Kernel for Relation
  Extraction}. In \bibinfo{booktitle}{\emph{Proceedings of the Conference on
  Human Language Technology and Empirical Methods in Natural Language
  Processing}} (Vancouver, British Columbia, Canada)
  \emph{(\bibinfo{series}{HLT '05})}. \bibinfo{publisher}{Association for
  Computational Linguistics}, \bibinfo{address}{USA},
  \bibinfo{pages}{724–731}.
\newblock
\urldef\tempurl%
\url{https://doi.org/10.3115/1220575.1220666}
\showDOI{\tempurl}


\bibitem[\protect\citeauthoryear{Ciampaglia}{Ciampaglia}{2018}]%
        {ciampaglia2018fighting}
\bibfield{author}{\bibinfo{person}{Giovanni~Luca Ciampaglia}.}
  \bibinfo{year}{2018}\natexlab{}.
\newblock \showarticletitle{Fighting fake news: a role for computational social
  science in the fight against digital misinformation}.
\newblock \bibinfo{journal}{\emph{Journal of Computational Social Science}}
  \bibinfo{volume}{1}, \bibinfo{number}{1} (\bibinfo{date}{29 Jan.}
  \bibinfo{year}{2018}), \bibinfo{pages}{147--153}.
\newblock
\showISSN{2432-2725}
\urldef\tempurl%
\url{https://doi.org/10.1007/s42001-017-0005-6}
\showDOI{\tempurl}


\bibitem[\protect\citeauthoryear{Ciampaglia, Shiralkar, Rocha, Bollen, Menczer,
  and Flammini}{Ciampaglia et~al\mbox{.}}{2015}]%
        {ciampaglia2015computational}
\bibfield{author}{\bibinfo{person}{Giovanni~Luca Ciampaglia},
  \bibinfo{person}{Prashant Shiralkar}, \bibinfo{person}{Luis~M. Rocha},
  \bibinfo{person}{Johan Bollen}, \bibinfo{person}{Filippo Menczer}, {and}
  \bibinfo{person}{Alessandro Flammini}.} \bibinfo{year}{2015}\natexlab{}.
\newblock \showarticletitle{Computational Fact Checking from Knowledge
  Networks}.
\newblock \bibinfo{journal}{\emph{PLOS ONE}} \bibinfo{volume}{10},
  \bibinfo{number}{6} (\bibinfo{date}{06} \bibinfo{year}{2015}),
  \bibinfo{pages}{1--13}.
\newblock
\urldef\tempurl%
\url{https://doi.org/10.1371/journal.pone.0128193}
\showDOI{\tempurl}


\bibitem[\protect\citeauthoryear{Dong, Gabrilovich, Heitz, Horn, Lao, Murphy,
  Strohmann, Sun, and Zhang}{Dong et~al\mbox{.}}{2014}]%
        {knowvault_Dong2014}
\bibfield{author}{\bibinfo{person}{Xin Dong}, \bibinfo{person}{Evgeniy
  Gabrilovich}, \bibinfo{person}{Geremy Heitz}, \bibinfo{person}{Wilko Horn},
  \bibinfo{person}{Ni Lao}, \bibinfo{person}{Kevin Murphy},
  \bibinfo{person}{Thomas Strohmann}, \bibinfo{person}{Shaohua Sun}, {and}
  \bibinfo{person}{Wei Zhang}.} \bibinfo{year}{2014}\natexlab{}.
\newblock \showarticletitle{{Knowledge vault: A web-scale approach to
  probabilistic knowledge fusion}}. In \bibinfo{booktitle}{\emph{Proceedings of
  the ACM SIGKDD International Conference on Knowledge Discovery and Data
  Mining}}. \bibinfo{publisher}{Association for Computing Machinery},
  \bibinfo{address}{New York, New York, USA}, \bibinfo{pages}{601--610}.
\newblock
\showISBNx{9781450329569}
\urldef\tempurl%
\url{https://doi.org/10.1145/2623330.2623623}
\showDOI{\tempurl}


\bibitem[\protect\citeauthoryear{Dong, Gabrilovich, Murphy, Dang, Horn,
  Lugaresi, Sun, and Zhang}{Dong et~al\mbox{.}}{2015}]%
        {dong_infotrust}
\bibfield{author}{\bibinfo{person}{Xin~Luna Dong}, \bibinfo{person}{Evgeniy
  Gabrilovich}, \bibinfo{person}{Kevin Murphy}, \bibinfo{person}{Van Dang},
  \bibinfo{person}{Wilko Horn}, \bibinfo{person}{Camillo Lugaresi},
  \bibinfo{person}{Shaohua Sun}, {and} \bibinfo{person}{Wei Zhang}.}
  \bibinfo{year}{2015}\natexlab{}.
\newblock \showarticletitle{Knowledge-Based Trust: Estimating the
  Trustworthiness of Web Sources}.
\newblock \bibinfo{journal}{\emph{Proc. VLDB Endow.}} \bibinfo{volume}{8},
  \bibinfo{number}{9} (\bibinfo{date}{May} \bibinfo{year}{2015}),
  \bibinfo{pages}{938–949}.
\newblock
\showISSN{2150-8097}
\urldef\tempurl%
\url{https://doi.org/10.14778/2777598.2777603}
\showDOI{\tempurl}


\bibitem[\protect\citeauthoryear{Fellbaum and Miller}{Fellbaum and
  Miller}{1998}]%
        {fellbaum1998wordnet}
\bibfield{author}{\bibinfo{person}{C. Fellbaum} {and} \bibinfo{person}{G.A.
  Miller}.} \bibinfo{year}{1998}\natexlab{}.
\newblock \bibinfo{booktitle}{\emph{WordNet: An Electronic Lexical Database}}.
\newblock \bibinfo{publisher}{MIT Press}, \bibinfo{address}{Cambridge, MA,
  USA}.
\newblock
\showISBNx{9780262061971}
\showLCCN{97048710}


\bibitem[\protect\citeauthoryear{Fridkin, Kenney, and Wintersieck}{Fridkin
  et~al\mbox{.}}{2015}]%
        {Fridkin2015}
\bibfield{author}{\bibinfo{person}{Kim Fridkin}, \bibinfo{person}{Patrick~J.
  Kenney}, {and} \bibinfo{person}{Amanda Wintersieck}.}
  \bibinfo{year}{2015}\natexlab{}.
\newblock \showarticletitle{{Liar, Liar, Pants on Fire: How Fact-Checking
  Influences Citizens' Reactions to Negative Advertising}}.
\newblock \bibinfo{journal}{\emph{Political Communication}}
  \bibinfo{volume}{32}, \bibinfo{number}{1} (\bibinfo{date}{Jan}
  \bibinfo{year}{2015}), \bibinfo{pages}{127--151}.
\newblock
\showISSN{10917675}
\urldef\tempurl%
\url{https://doi.org/10.1080/10584609.2014.914613}
\showDOI{\tempurl}


\bibitem[\protect\citeauthoryear{Gal{\'{a}}rraga, Teflioudi, Hose, and
  Suchanek}{Gal{\'{a}}rraga et~al\mbox{.}}{2013}]%
        {Galarraga2013}
\bibfield{author}{\bibinfo{person}{Luis~Antonio Gal{\'{a}}rraga},
  \bibinfo{person}{Christina Teflioudi}, \bibinfo{person}{Katja Hose}, {and}
  \bibinfo{person}{Fabian Suchanek}.} \bibinfo{year}{2013}\natexlab{}.
\newblock \showarticletitle{{AMIE: association rule mining under incomplete
  evidence in ontological knowledge bases}}. In
  \bibinfo{booktitle}{\emph{Proceedings of the 22nd international conference on
  World Wide Web - WWW '13}}. \bibinfo{publisher}{ACM Press},
  \bibinfo{address}{New York, New York, USA}, \bibinfo{pages}{413--422}.
\newblock
\showISBNx{9781450320351}
\urldef\tempurl%
\url{https://doi.org/10.1145/2488388.2488425}
\showDOI{\tempurl}


\bibitem[\protect\citeauthoryear{Gangemi, Presutti, {Reforgiato Recupero},
  Nuzzolese, Draicchio, and Mongiov{\`{i}}}{Gangemi et~al\mbox{.}}{2017}]%
        {Gangemi2017}
\bibfield{author}{\bibinfo{person}{Aldo Gangemi}, \bibinfo{person}{Valentina
  Presutti}, \bibinfo{person}{Diego {Reforgiato Recupero}},
  \bibinfo{person}{Andrea~Giovanni Nuzzolese}, \bibinfo{person}{Francesco
  Draicchio}, {and} \bibinfo{person}{Misael Mongiov{\`{i}}}.}
  \bibinfo{year}{2017}\natexlab{}.
\newblock \showarticletitle{{Semantic Web Machine Reading with FRED}}.
\newblock \bibinfo{journal}{\emph{Semantic Web}} \bibinfo{volume}{8},
  \bibinfo{number}{6} (\bibinfo{year}{2017}), \bibinfo{pages}{873--893}.
\newblock
\showISSN{22104968}
\urldef\tempurl%
\url{https://doi.org/10.3233/SW-160240}
\showDOI{\tempurl}


\bibitem[\protect\citeauthoryear{Gardner and Mitchell}{Gardner and
  Mitchell}{2015}]%
        {gardner2015efficient}
\bibfield{author}{\bibinfo{person}{Matt Gardner} {and} \bibinfo{person}{Tom
  Mitchell}.} \bibinfo{year}{2015}\natexlab{}.
\newblock \showarticletitle{Efficient and Expressive Knowledge Base Completion
  Using Subgraph Feature Extraction}. In \bibinfo{booktitle}{\emph{Proceedings
  of the 2015 Conference on Empirical Methods in Natural Language Processing}}.
  \bibinfo{publisher}{Association for Computational Linguistics},
  \bibinfo{address}{Lisbon, Portugal}, \bibinfo{pages}{1488--1498}.
\newblock
\urldef\tempurl%
\url{https://doi.org/10.18653/v1/D15-1173}
\showDOI{\tempurl}


\bibitem[\protect\citeauthoryear{Grinberg, Joseph, Friedland, Swire-Thompson,
  and Lazer}{Grinberg et~al\mbox{.}}{2019}]%
        {grinberg2019fake}
\bibfield{author}{\bibinfo{person}{Nir Grinberg}, \bibinfo{person}{Kenneth
  Joseph}, \bibinfo{person}{Lisa Friedland}, \bibinfo{person}{Briony
  Swire-Thompson}, {and} \bibinfo{person}{David Lazer}.}
  \bibinfo{year}{2019}\natexlab{}.
\newblock \showarticletitle{Fake news on Twitter during the 2016 US
  presidential election}.
\newblock \bibinfo{journal}{\emph{Science}} \bibinfo{volume}{363},
  \bibinfo{number}{6425} (\bibinfo{year}{2019}), \bibinfo{pages}{374--378}.
\newblock


\bibitem[\protect\citeauthoryear{Grover and Leskovec}{Grover and
  Leskovec}{2016}]%
        {Grover2016}
\bibfield{author}{\bibinfo{person}{Aditya Grover} {and} \bibinfo{person}{Jure
  Leskovec}.} \bibinfo{year}{2016}\natexlab{}.
\newblock \showarticletitle{{Node2vec: Scalable feature learning for
  networks}}.
\newblock \bibinfo{journal}{\emph{Proceedings of the ACM SIGKDD International
  Conference on Knowledge Discovery and Data Mining}}
  \bibinfo{volume}{13-17-Augu} (\bibinfo{year}{2016}),
  \bibinfo{pages}{855--864}.
\newblock
\showISBNx{9781450342322}
\urldef\tempurl%
\url{https://doi.org/10.1145/2939672.2939754}
\showDOI{\tempurl}
\showeprint[arxiv]{1607.00653}


\bibitem[\protect\citeauthoryear{Guess, Nyhan, and Reifler}{Guess
  et~al\mbox{.}}{2020}]%
        {guess2020exposure}
\bibfield{author}{\bibinfo{person}{Andrew~M Guess}, \bibinfo{person}{Brendan
  Nyhan}, {and} \bibinfo{person}{Jason Reifler}.}
  \bibinfo{year}{2020}\natexlab{}.
\newblock \showarticletitle{Exposure to untrustworthy websites in the 2016 US
  election}.
\newblock \bibinfo{journal}{\emph{Nature human behaviour}} \bibinfo{volume}{4},
  \bibinfo{number}{5} (\bibinfo{year}{2020}), \bibinfo{pages}{472--480}.
\newblock


\bibitem[\protect\citeauthoryear{Guha, Brickley, and Macbeth}{Guha
  et~al\mbox{.}}{2016}]%
        {guha2016schema.org}
\bibfield{author}{\bibinfo{person}{R.~V. Guha}, \bibinfo{person}{Dan Brickley},
  {and} \bibinfo{person}{Steve Macbeth}.} \bibinfo{year}{2016}\natexlab{}.
\newblock \showarticletitle{Schema.Org: Evolution of Structured Data on the
  Web}.
\newblock \bibinfo{journal}{\emph{Commun. ACM}} \bibinfo{volume}{59},
  \bibinfo{number}{2} (\bibinfo{date}{Jan.} \bibinfo{year}{2016}),
  \bibinfo{pages}{44–51}.
\newblock
\showISSN{0001-0782}
\urldef\tempurl%
\url{https://doi.org/10.1145/2844544}
\showDOI{\tempurl}


\bibitem[\protect\citeauthoryear{Hamilton, Ying, and Leskovec}{Hamilton
  et~al\mbox{.}}{2017}]%
        {hamilton-graphsage}
\bibfield{author}{\bibinfo{person}{William~L. Hamilton}, \bibinfo{person}{Rex
  Ying}, {and} \bibinfo{person}{Jure Leskovec}.}
  \bibinfo{year}{2017}\natexlab{}.
\newblock \showarticletitle{Inductive Representation Learning on Large Graphs}.
  In \bibinfo{booktitle}{\emph{Proceedings of the 31st International Conference
  on Neural Information Processing Systems}} (Long Beach, California, USA)
  \emph{(\bibinfo{series}{NIPS'17})}. \bibinfo{publisher}{Curran Associates
  Inc.}, \bibinfo{address}{Red Hook, NY, USA}, \bibinfo{pages}{1025–1035}.
\newblock
\showISBNx{9781510860964}


\bibitem[\protect\citeauthoryear{Hassan, Zhang, Arslan, Caraballo, Jimenez,
  Gawsane, Hasan, Joseph, Kulkarni, Nayak, Sable, Li, and Tremayne}{Hassan
  et~al\mbox{.}}{2017}]%
        {Hassan2017}
\bibfield{author}{\bibinfo{person}{Naeemul Hassan}, \bibinfo{person}{Gensheng
  Zhang}, \bibinfo{person}{Fatma Arslan}, \bibinfo{person}{Josue Caraballo},
  \bibinfo{person}{Damian Jimenez}, \bibinfo{person}{Siddhant Gawsane},
  \bibinfo{person}{Shohedul Hasan}, \bibinfo{person}{Minumol Joseph},
  \bibinfo{person}{Aaditya Kulkarni}, \bibinfo{person}{Anil~Kumar Nayak},
  \bibinfo{person}{Vikas Sable}, \bibinfo{person}{Chengkai Li}, {and}
  \bibinfo{person}{Mark Tremayne}.} \bibinfo{year}{2017}\natexlab{}.
\newblock \showarticletitle{{Claim buster: The firstever endtoend factchecking
  system}}.
\newblock \bibinfo{journal}{\emph{Proceedings of the VLDB Endowment}}
  \bibinfo{volume}{10}, \bibinfo{number}{12} (\bibinfo{year}{2017}),
  \bibinfo{pages}{1945--1948}.
\newblock
\showISSN{21508097}
\urldef\tempurl%
\url{https://doi.org/10.14778/3137765.3137815}
\showDOI{\tempurl}


\bibitem[\protect\citeauthoryear{Huynh and Papotti}{Huynh and Papotti}{2019}]%
        {Huynh2019}
\bibfield{author}{\bibinfo{person}{Viet~Phi Huynh} {and} \bibinfo{person}{Paolo
  Papotti}.} \bibinfo{year}{2019}\natexlab{}.
\newblock \showarticletitle{{A benchmark for fact checking algorithms built on
  knowledge bases}}. In \bibinfo{booktitle}{\emph{International Conference on
  Information and Knowledge Management, Proceedings}},
  Vol.~\bibinfo{volume}{10}. \bibinfo{publisher}{Association for Computing
  Machinery}, \bibinfo{address}{New York, NY, USA}, \bibinfo{pages}{689--698}.
\newblock
\showISBNx{9781450369763}
\urldef\tempurl%
\url{https://doi.org/10.1145/3357384.3358036}
\showDOI{\tempurl}


\bibitem[\protect\citeauthoryear{Kale}{Kale}{2018}]%
        {kale2018no}
\bibfield{author}{\bibinfo{person}{Krutika Kale}.}
  \bibinfo{year}{2018}\natexlab{}.
\newblock \bibinfo{title}{No, {T}ej {P}ratap {Y}adav Did Not Receive A
  Doctorate From {T}akshashila {U}niversity}.
\newblock \bibinfo{howpublished}{Online at
  \url{https://www.boomlive.in/no-lalus-son-tej-pratap-did-not-receive-a-doctorate-from-takshashila-university/}}.
\newblock
\newblock
\shownote{Last accessed 2021-02-21.}


\bibitem[\protect\citeauthoryear{Lai, Leung, and Leung}{Lai
  et~al\mbox{.}}{2018}]%
        {lai-etal-2018-sunnynlp}
\bibfield{author}{\bibinfo{person}{Sunny Lai}, \bibinfo{person}{Kwong~Sak
  Leung}, {and} \bibinfo{person}{Yee Leung}.} \bibinfo{year}{2018}\natexlab{}.
\newblock \showarticletitle{{SUNNYNLP} at {S}em{E}val-2018 Task 10: A
  Support-Vector-Machine-Based Method for Detecting Semantic Difference using
  Taxonomy and Word Embedding Features}. In
  \bibinfo{booktitle}{\emph{Proceedings of The 12th International Workshop on
  Semantic Evaluation}}. \bibinfo{publisher}{Association for Computational
  Linguistics}, \bibinfo{address}{New Orleans, Louisiana},
  \bibinfo{pages}{741--746}.
\newblock
\urldef\tempurl%
\url{https://doi.org/10.18653/v1/S18-1118}
\showDOI{\tempurl}


\bibitem[\protect\citeauthoryear{Lao and Cohen}{Lao and Cohen}{2010}]%
        {Lao2010}
\bibfield{author}{\bibinfo{person}{Ni Lao} {and} \bibinfo{person}{William~W.
  Cohen}.} \bibinfo{year}{2010}\natexlab{}.
\newblock \showarticletitle{{Relational retrieval using a combination of
  path-constrained random walks}}.
\newblock \bibinfo{journal}{\emph{Machine Learning}} \bibinfo{volume}{81},
  \bibinfo{number}{1} (\bibinfo{year}{2010}), \bibinfo{pages}{53--67}.
\newblock
\showISSN{08856125}
\urldef\tempurl%
\url{https://doi.org/10.1007/s10994-010-5205-8}
\showDOI{\tempurl}


\bibitem[\protect\citeauthoryear{Lao, Mitchell, and Cohen}{Lao
  et~al\mbox{.}}{2011}]%
        {lao2011random}
\bibfield{author}{\bibinfo{person}{Ni Lao}, \bibinfo{person}{Tom Mitchell},
  {and} \bibinfo{person}{William~W. Cohen}.} \bibinfo{year}{2011}\natexlab{}.
\newblock \showarticletitle{Random Walk Inference and Learning in A Large Scale
  Knowledge Base}. In \bibinfo{booktitle}{\emph{Proceedings of the 2011
  Conference on Empirical Methods in Natural Language Processing}}.
  \bibinfo{publisher}{Association for Computational Linguistics},
  \bibinfo{address}{Edinburgh, Scotland, UK.}, \bibinfo{pages}{529--539}.
\newblock


\bibitem[\protect\citeauthoryear{Lazer, Baum, Benkler, Berinsky, Greenhill,
  Menczer, Metzger, Nyhan, Pennycook, Rothschild, et~al\mbox{.}}{Lazer
  et~al\mbox{.}}{2018}]%
        {lazer2018science}
\bibfield{author}{\bibinfo{person}{David~MJ Lazer}, \bibinfo{person}{Matthew~A
  Baum}, \bibinfo{person}{Yochai Benkler}, \bibinfo{person}{Adam~J Berinsky},
  \bibinfo{person}{Kelly~M Greenhill}, \bibinfo{person}{Filippo Menczer},
  \bibinfo{person}{Miriam~J Metzger}, \bibinfo{person}{Brendan Nyhan},
  \bibinfo{person}{Gordon Pennycook}, \bibinfo{person}{David Rothschild},
  {et~al\mbox{.}}} \bibinfo{year}{2018}\natexlab{}.
\newblock \showarticletitle{The science of fake news}.
\newblock \bibinfo{journal}{\emph{Science}} \bibinfo{volume}{359},
  \bibinfo{number}{6380} (\bibinfo{year}{2018}), \bibinfo{pages}{1094--1096}.
\newblock


\bibitem[\protect\citeauthoryear{Le and Mikolov}{Le and Mikolov}{2014}]%
        {le2014distributed}
\bibfield{author}{\bibinfo{person}{Quoc Le} {and} \bibinfo{person}{Tomas
  Mikolov}.} \bibinfo{year}{2014}\natexlab{}.
\newblock \showarticletitle{Distributed Representations of Sentences and
  Documents}. In \bibinfo{booktitle}{\emph{Proceedings of the 31st
  International Conference on International Conference on Machine Learning -
  Volume 32}} \emph{(\bibinfo{series}{ICML'14})}.
  \bibinfo{publisher}{JMLR.org}, \bibinfo{address}{Beijing, China},
  \bibinfo{pages}{II–1188–II–1196}.
\newblock


\bibitem[\protect\citeauthoryear{Lewandowsky, Ecker, Seifert, Schwarz, and
  Cook}{Lewandowsky et~al\mbox{.}}{2012}]%
        {lewandowsky2012misinformation}
\bibfield{author}{\bibinfo{person}{Stephan Lewandowsky},
  \bibinfo{person}{Ullrich K.~H. Ecker}, \bibinfo{person}{Colleen~M. Seifert},
  \bibinfo{person}{Norbert Schwarz}, {and} \bibinfo{person}{John Cook}.}
  \bibinfo{year}{2012}\natexlab{}.
\newblock \showarticletitle{Misinformation and Its Correction: Continued
  Influence and Successful Debiasing}.
\newblock \bibinfo{journal}{\emph{Psychological Science in the Public
  Interest}} \bibinfo{volume}{13}, \bibinfo{number}{3} (\bibinfo{year}{2012}),
  \bibinfo{pages}{106--131}.
\newblock
\urldef\tempurl%
\url{https://doi.org/10.1177/1529100612451018}
\showDOI{\tempurl}


\bibitem[\protect\citeauthoryear{Liben-Nowell and Kleinberg}{Liben-Nowell and
  Kleinberg}{2007}]%
        {liben2007link}
\bibfield{author}{\bibinfo{person}{David Liben-Nowell} {and}
  \bibinfo{person}{Jon Kleinberg}.} \bibinfo{year}{2007}\natexlab{}.
\newblock \showarticletitle{The Link-Prediction Problem for Social Networks}.
\newblock \bibinfo{journal}{\emph{Journal of the American society for
  Information Science and Technology}} \bibinfo{volume}{58},
  \bibinfo{number}{7} (\bibinfo{year}{2007}), \bibinfo{pages}{1019--1031}.
\newblock


\bibitem[\protect\citeauthoryear{Liu, Nourbakhsh, Li, Fang, and Shah}{Liu
  et~al\mbox{.}}{2015}]%
        {liu2015realtime}
\bibfield{author}{\bibinfo{person}{Xiaomo Liu}, \bibinfo{person}{Armineh
  Nourbakhsh}, \bibinfo{person}{Quanzhi Li}, \bibinfo{person}{Rui Fang}, {and}
  \bibinfo{person}{Sameena Shah}.} \bibinfo{year}{2015}\natexlab{}.
\newblock \showarticletitle{Real-Time Rumor Debunking on Twitter}. In
  \bibinfo{booktitle}{\emph{Proceedings of the 24th ACM International on
  Conference on Information and Knowledge Management}} (Melbourne, Australia)
  \emph{(\bibinfo{series}{CIKM '15})}. \bibinfo{publisher}{Association for
  Computing Machinery}, \bibinfo{address}{New York, NY, USA},
  \bibinfo{pages}{1867--1870}.
\newblock
\showISBNx{9781450337946}
\urldef\tempurl%
\url{https://doi.org/10.1145/2806416.2806651}
\showDOI{\tempurl}


\bibitem[\protect\citeauthoryear{Mitchell, Cohen, Hruschka, Talukdar, Yang,
  Betteridge, Carlson, Dalvi, Gardner, Kisiel, Krishnamurthy, Lao, Mazaitis,
  Mohamed, Nakashole, Platanios, Ritter, Samadi, Settles, Wang, Wijaya, Gupta,
  Chen, Saparov, Greaves, and Welling}{Mitchell et~al\mbox{.}}{2018}]%
        {mitchell2018never}
\bibfield{author}{\bibinfo{person}{T. Mitchell}, \bibinfo{person}{W. Cohen},
  \bibinfo{person}{E. Hruschka}, \bibinfo{person}{P. Talukdar},
  \bibinfo{person}{B. Yang}, \bibinfo{person}{J. Betteridge},
  \bibinfo{person}{A. Carlson}, \bibinfo{person}{B. Dalvi}, \bibinfo{person}{M.
  Gardner}, \bibinfo{person}{B. Kisiel}, \bibinfo{person}{J. Krishnamurthy},
  \bibinfo{person}{N. Lao}, \bibinfo{person}{K. Mazaitis}, \bibinfo{person}{T.
  Mohamed}, \bibinfo{person}{N. Nakashole}, \bibinfo{person}{E. Platanios},
  \bibinfo{person}{A. Ritter}, \bibinfo{person}{M. Samadi}, \bibinfo{person}{B.
  Settles}, \bibinfo{person}{R. Wang}, \bibinfo{person}{D. Wijaya},
  \bibinfo{person}{A. Gupta}, \bibinfo{person}{X. Chen}, \bibinfo{person}{A.
  Saparov}, \bibinfo{person}{M. Greaves}, {and} \bibinfo{person}{J. Welling}.}
  \bibinfo{year}{2018}\natexlab{}.
\newblock \showarticletitle{Never-Ending Learning}.
\newblock \bibinfo{journal}{\emph{Commun. ACM}} \bibinfo{volume}{61},
  \bibinfo{number}{5} (\bibinfo{date}{April} \bibinfo{year}{2018}),
  \bibinfo{pages}{103–115}.
\newblock
\showISSN{0001-0782}
\urldef\tempurl%
\url{https://doi.org/10.1145/3191513}
\showDOI{\tempurl}


\bibitem[\protect\citeauthoryear{Nyhan and Reifler}{Nyhan and Reifler}{2015}]%
        {nyhan2015effect}
\bibfield{author}{\bibinfo{person}{Brendan Nyhan} {and} \bibinfo{person}{Jason
  Reifler}.} \bibinfo{year}{2015}\natexlab{}.
\newblock \showarticletitle{The Effect of Fact-Checking on Elites: A Field
  Experiment on U.S. State Legislators}.
\newblock \bibinfo{journal}{\emph{American Journal of Political Science}}
  \bibinfo{volume}{59}, \bibinfo{number}{3} (\bibinfo{year}{2015}),
  \bibinfo{pages}{628--640}.
\newblock
\urldef\tempurl%
\url{https://doi.org/10.1111/ajps.12162}
\showDOI{\tempurl}


\bibitem[\protect\citeauthoryear{Orr}{Orr}{2013}]%
        {orr_2013}
\bibfield{author}{\bibinfo{person}{Dave Orr}.} \bibinfo{year}{2013}\natexlab{}.
\newblock \bibinfo{title}{50,000 Lessons on How to Read: a Relation Extraction
  Corpus}.
\newblock
\newblock
\urldef\tempurl%
\url{https://ai.googleblog.com/2013/04/50000-lessons-on-how-to-read-relation.html}
\showURL{%
\tempurl}


\bibitem[\protect\citeauthoryear{Ortona, Meduri, and Papotti}{Ortona
  et~al\mbox{.}}{2018}]%
        {ortona2018robust}
\bibfield{author}{\bibinfo{person}{Stefano Ortona},
  \bibinfo{person}{Venkata~Vamsikrishna Meduri}, {and} \bibinfo{person}{Paolo
  Papotti}.} \bibinfo{year}{2018}\natexlab{}.
\newblock \showarticletitle{Robust discovery of positive and negative rules in
  knowledge bases}. In \bibinfo{booktitle}{\emph{2018 IEEE 34th International
  Conference on Data Engineering (ICDE)}} (Paris, France). IEEE,
  \bibinfo{publisher}{IEEE}, \bibinfo{address}{Piscataway, NJ, USA},
  \bibinfo{pages}{1168--1179}.
\newblock


\bibitem[\protect\citeauthoryear{Paulheim and Ponzetto}{Paulheim and
  Ponzetto}{2013}]%
        {paulheim_wikiaugment}
\bibfield{author}{\bibinfo{person}{Heiko Paulheim} {and}
  \bibinfo{person}{Simone~Paolo Ponzetto}.} \bibinfo{year}{2013}\natexlab{}.
\newblock \showarticletitle{Extending DBpedia with Wikipedia List Pages}. In
  \bibinfo{booktitle}{\emph{Proceedings of the 2013th International Conference
  on NLP \& DBpedia - Volume 1064}} (Sydney, Australia)
  \emph{(\bibinfo{series}{NLP-DBPEDIA'13})}. \bibinfo{publisher}{CEUR-WS.org},
  \bibinfo{address}{Aachen, DEU}, \bibinfo{pages}{85–90}.
\newblock


\bibitem[\protect\citeauthoryear{Rashkin, Choi, Jang, Volkova, and
  Choi}{Rashkin et~al\mbox{.}}{2017}]%
        {rashkin-etal-2017-truth}
\bibfield{author}{\bibinfo{person}{Hannah Rashkin}, \bibinfo{person}{Eunsol
  Choi}, \bibinfo{person}{Jin~Yea Jang}, \bibinfo{person}{Svitlana Volkova},
  {and} \bibinfo{person}{Yejin Choi}.} \bibinfo{year}{2017}\natexlab{}.
\newblock \showarticletitle{Truth of Varying Shades: Analyzing Language in Fake
  News and Political Fact-Checking}. In \bibinfo{booktitle}{\emph{Proceedings
  of the 2017 Conference on Empirical Methods in Natural Language Processing}}.
  \bibinfo{publisher}{Association for Computational Linguistics},
  \bibinfo{address}{Copenhagen, Denmark}, \bibinfo{pages}{2931--2937}.
\newblock
\urldef\tempurl%
\url{https://doi.org/10.18653/v1/D17-1317}
\showDOI{\tempurl}


\bibitem[\protect\citeauthoryear{Richards and Mooney}{Richards and
  Mooney}{1992}]%
        {bradley1992learning}
\bibfield{author}{\bibinfo{person}{Bradley~L. Richards} {and}
  \bibinfo{person}{Raymond~J. Mooney}.} \bibinfo{year}{1992}\natexlab{}.
\newblock \showarticletitle{Learning Relations by Pathfinding}. In
  \bibinfo{booktitle}{\emph{Proceedings of the Tenth National Conference on
  Artificial Intelligence}} (San Jose, California)
  \emph{(\bibinfo{series}{AAAI'92})}. \bibinfo{publisher}{AAAI Press},
  \bibinfo{address}{Palo Alto, CA, USA}, \bibinfo{pages}{50--55}.
\newblock
\showISBNx{0262510634}


\bibitem[\protect\citeauthoryear{Rubin, Chen, and Conroy}{Rubin
  et~al\mbox{.}}{2015}]%
        {doi:10.1002/pra2.2015.145052010083}
\bibfield{author}{\bibinfo{person}{Victoria~L. Rubin}, \bibinfo{person}{Yimin
  Chen}, {and} \bibinfo{person}{Nadia~K. Conroy}.}
  \bibinfo{year}{2015}\natexlab{}.
\newblock \showarticletitle{Deception detection for news: Three types of
  fakes}.
\newblock \bibinfo{journal}{\emph{Proceedings of the Association for
  Information Science and Technology}} \bibinfo{volume}{52},
  \bibinfo{number}{1} (\bibinfo{year}{2015}), \bibinfo{pages}{1--4}.
\newblock
\urldef\tempurl%
\url{https://doi.org/10.1002/pra2.2015.145052010083}
\showDOI{\tempurl}


\bibitem[\protect\citeauthoryear{Rubin and Vashchilko}{Rubin and
  Vashchilko}{2012}]%
        {rubin-vashchilko-2012-identification}
\bibfield{author}{\bibinfo{person}{Victoria~L. Rubin} {and}
  \bibinfo{person}{Tatiana Vashchilko}.} \bibinfo{year}{2012}\natexlab{}.
\newblock \showarticletitle{Identification of Truth and Deception in Text:
  Application of Vector Space Model to {R}hetorical {S}tructure {T}heory}. In
  \bibinfo{booktitle}{\emph{Proceedings of the Workshop on Computational
  Approaches to Deception Detection}}. \bibinfo{publisher}{Association for
  Computational Linguistics}, \bibinfo{address}{Avignon, France},
  \bibinfo{pages}{97--106}.
\newblock


\bibitem[\protect\citeauthoryear{schema.org}{schema.org}{2020}]%
        {schema.org}
\bibfield{author}{\bibinfo{person}{schema.org}.}
  \bibinfo{year}{2020}\natexlab{}.
\newblock \bibinfo{title}{ClaimReview schema}.
\newblock
\newblock
\urldef\tempurl%
\url{https://schema.org/ClaimReview}
\showURL{%
\tempurl}


\bibitem[\protect\citeauthoryear{Shao, Ciampaglia, Flammini, and Menczer}{Shao
  et~al\mbox{.}}{2016}]%
        {shao2016hoaxy}
\bibfield{author}{\bibinfo{person}{Chengcheng Shao},
  \bibinfo{person}{Giovanni~Luca Ciampaglia}, \bibinfo{person}{Alessandro
  Flammini}, {and} \bibinfo{person}{Filippo Menczer}.}
  \bibinfo{year}{2016}\natexlab{}.
\newblock \showarticletitle{Hoaxy: A Platform for Tracking Online
  Misinformation}. In \bibinfo{booktitle}{\emph{Proceedings of the
  25\textsuperscript{th} International Conference Companion on World Wide Web}}
  (Montr\'{e}al, Qu\'{e}bec, Canada) \emph{(\bibinfo{series}{WWW '16
  Companion})}. \bibinfo{publisher}{International World Wide Web Conferences
  Steering Committee}, \bibinfo{address}{Republic and Canton of Geneva,
  Switzerland}, \bibinfo{pages}{745--750}.
\newblock
\showISBNx{978-1-4503-4144-8}
\urldef\tempurl%
\url{https://doi.org/10.1145/2872518.2890098}
\showDOI{\tempurl}


\bibitem[\protect\citeauthoryear{Shao, Ciampaglia, Varol, Yang, Flammini, and
  Menczer}{Shao et~al\mbox{.}}{2018}]%
        {shao2018spread}
\bibfield{author}{\bibinfo{person}{Chengcheng Shao},
  \bibinfo{person}{Giovanni~Luca Ciampaglia}, \bibinfo{person}{Onur Varol},
  \bibinfo{person}{Kai-Cheng Yang}, \bibinfo{person}{Alessandro Flammini},
  {and} \bibinfo{person}{Filippo Menczer}.} \bibinfo{year}{2018}\natexlab{}.
\newblock \showarticletitle{The spread of low-credibility content by social
  bots}.
\newblock \bibinfo{journal}{\emph{Nature communications}} \bibinfo{volume}{9},
  \bibinfo{number}{1} (\bibinfo{year}{2018}), \bibinfo{pages}{1--9}.
\newblock


\bibitem[\protect\citeauthoryear{Shi and Weninger}{Shi and Weninger}{2016}]%
        {Shi2016}
\bibfield{author}{\bibinfo{person}{Baoxu Shi} {and} \bibinfo{person}{Tim
  Weninger}.} \bibinfo{year}{2016}\natexlab{}.
\newblock \showarticletitle{{Discriminative predicate path mining for fact
  checking in knowledge graphs}}.
\newblock \bibinfo{journal}{\emph{Knowledge-Based Systems}}
  \bibinfo{volume}{104} (\bibinfo{date}{Jul} \bibinfo{year}{2016}),
  \bibinfo{pages}{123--133}.
\newblock
\showISSN{09507051}
\urldef\tempurl%
\url{https://doi.org/10.1016/j.knosys.2016.04.015}
\showDOI{\tempurl}
\showeprint[arxiv]{1510.05911}


\bibitem[\protect\citeauthoryear{Shiralkar, Flammini, Menczer, and
  Ciampaglia}{Shiralkar et~al\mbox{.}}{2017}]%
        {shiralkar2017finding}
\bibfield{author}{\bibinfo{person}{Prashant Shiralkar},
  \bibinfo{person}{Alessandro Flammini}, \bibinfo{person}{Filippo Menczer},
  {and} \bibinfo{person}{Giovanni~Luca Ciampaglia}.}
  \bibinfo{year}{2017}\natexlab{}.
\newblock \showarticletitle{Finding Streams in Knowledge Graphs to Support Fact
  Checking}. In \bibinfo{booktitle}{\emph{2017 IEEE International Conference on
  Data Mining (ICDM)}} (New Orleans, Louisiana, USA).
  \bibinfo{publisher}{IEEE}, \bibinfo{address}{Piscataway, NJ},
  \bibinfo{pages}{859--864}.
\newblock
\showISBNx{2374-8486}
\urldef\tempurl%
\url{https://doi.org/10.1109/ICDM.2017.105}
\showDOI{\tempurl}
\showeprint[arxiv]{1708.07239}~[cs.AI]
\newblock
\shownote{Extended Version.}


\bibitem[\protect\citeauthoryear{Shu, Mahudeswaran, Wang, and Liu}{Shu
  et~al\mbox{.}}{2020}]%
        {shu2020hierarchical}
\bibfield{author}{\bibinfo{person}{Kai Shu}, \bibinfo{person}{Deepak
  Mahudeswaran}, \bibinfo{person}{Suhang Wang}, {and} \bibinfo{person}{Huan
  Liu}.} \bibinfo{year}{2020}\natexlab{}.
\newblock \showarticletitle{Hierarchical propagation networks for fake news
  detection: Investigation and exploitation}. In
  \bibinfo{booktitle}{\emph{Proceedings of the International AAAI Conference on
  Web and Social Media}}, Vol.~\bibinfo{volume}{14}. \bibinfo{publisher}{AAAI},
  \bibinfo{address}{Palo Alto, CA, USA}, \bibinfo{pages}{626--637}.
\newblock


\bibitem[\protect\citeauthoryear{Shu, Wang, and Liu}{Shu et~al\mbox{.}}{2018}]%
        {Shu2018}
\bibfield{author}{\bibinfo{person}{Kai Shu}, \bibinfo{person}{Suhang Wang},
  {and} \bibinfo{person}{Huan Liu}.} \bibinfo{year}{2018}\natexlab{}.
\newblock \showarticletitle{Understanding User Profiles on Social Media for
  Fake News Detection}. In \bibinfo{booktitle}{\emph{IEEE 1st Conference on
  Multimedia Information Processing and Retrieval}}
  \emph{(\bibinfo{series}{MIPR 2018})}. \bibinfo{publisher}{IEEE},
  \bibinfo{address}{Piscataway, NJ, USA}, \bibinfo{pages}{430--435}.
\newblock
\showISBNx{9781538618578}
\urldef\tempurl%
\url{https://doi.org/10.1109/MIPR.2018.00092}
\showDOI{\tempurl}


\bibitem[\protect\citeauthoryear{Silverman}{Silverman}{2014}]%
        {silverman2014verification}
\bibfield{editor}{\bibinfo{person}{Craig Silverman}} (Ed.).
  \bibinfo{year}{2014}\natexlab{}.
\newblock \bibinfo{booktitle}{\emph{Verification Handbook}}.
\newblock \bibinfo{publisher}{European Journalism Center},
  \bibinfo{address}{Maastricht, the Netherlands}.
\newblock


\bibitem[\protect\citeauthoryear{Singh, Riedel, Martin, Zheng, and
  McCallum}{Singh et~al\mbox{.}}{2013}]%
        {singh2013joint}
\bibfield{author}{\bibinfo{person}{Sameer Singh}, \bibinfo{person}{Sebastian
  Riedel}, \bibinfo{person}{Brian Martin}, \bibinfo{person}{Jiaping Zheng},
  {and} \bibinfo{person}{Andrew McCallum}.} \bibinfo{year}{2013}\natexlab{}.
\newblock \showarticletitle{Joint Inference of Entities, Relations, and
  Coreference}. In \bibinfo{booktitle}{\emph{Proceedings of the 2013 Workshop
  on Automated Knowledge Base Construction}} (San Francisco, California, USA)
  \emph{(\bibinfo{series}{AKBC '13})}. \bibinfo{publisher}{Association for
  Computing Machinery}, \bibinfo{address}{New York, NY, USA},
  \bibinfo{pages}{1–6}.
\newblock
\showISBNx{9781450324113}
\urldef\tempurl%
\url{https://doi.org/10.1145/2509558.2509559}
\showDOI{\tempurl}


\bibitem[\protect\citeauthoryear{Socher, Chen, Manning, and Ng}{Socher
  et~al\mbox{.}}{2013}]%
        {socher_nn_kbaugment}
\bibfield{author}{\bibinfo{person}{Richard Socher}, \bibinfo{person}{Danqi
  Chen}, \bibinfo{person}{Christopher~D Manning}, {and} \bibinfo{person}{Andrew
  Ng}.} \bibinfo{year}{2013}\natexlab{}.
\newblock \showarticletitle{Reasoning With Neural Tensor Networks for Knowledge
  Base Completion}. In \bibinfo{booktitle}{\emph{Advances in Neural Information
  Processing Systems}}, \bibfield{editor}{\bibinfo{person}{C.~J.~C. Burges},
  \bibinfo{person}{L.~Bottou}, \bibinfo{person}{M.~Welling},
  \bibinfo{person}{Z.~Ghahramani}, {and} \bibinfo{person}{K.~Q. Weinberger}}
  (Eds.), Vol.~\bibinfo{volume}{26}. \bibinfo{publisher}{Curran Associates,
  Inc.}, \bibinfo{address}{57 Morehouse Lane, Red Hook, NY, United States},
  \bibinfo{pages}{926--934}.
\newblock
\urldef\tempurl%
\url{https://proceedings.neurips.cc/paper/2013/file/b337e84de8752b27eda3a12363109e80-Paper.pdf}
\showURL{%
\tempurl}


\bibitem[\protect\citeauthoryear{Thorne, Vlachos, Christodoulopoulos, and
  Mittal}{Thorne et~al\mbox{.}}{2018}]%
        {Thorne2018}
\bibfield{author}{\bibinfo{person}{James Thorne}, \bibinfo{person}{Andreas
  Vlachos}, \bibinfo{person}{Christos Christodoulopoulos}, {and}
  \bibinfo{person}{Arpit Mittal}.} \bibinfo{year}{2018}\natexlab{}.
\newblock \showarticletitle{{FEVER: a Large-scale Dataset for Fact Extraction
  and VERification}}. In \bibinfo{booktitle}{\emph{Proceedings of the 2018
  Conference of the North American Chapter of the Association for Computational
  Linguistics: Human Language Technologies, Volume 1 (Long Papers)}}.
  \bibinfo{publisher}{Association for Computational Linguistics},
  \bibinfo{address}{Stroudsburg, PA, USA}, \bibinfo{pages}{809--819}.
\newblock
\urldef\tempurl%
\url{https://doi.org/10.18653/v1/N18-1074}
\showDOI{\tempurl}


\bibitem[\protect\citeauthoryear{Tschiatschek, Singla, Gomez~Rodriguez,
  Merchant, and Krause}{Tschiatschek et~al\mbox{.}}{2018}]%
        {10.1145/3184558.3188722}
\bibfield{author}{\bibinfo{person}{Sebastian Tschiatschek},
  \bibinfo{person}{Adish Singla}, \bibinfo{person}{Manuel Gomez~Rodriguez},
  \bibinfo{person}{Arpit Merchant}, {and} \bibinfo{person}{Andreas Krause}.}
  \bibinfo{year}{2018}\natexlab{}.
\newblock \showarticletitle{Fake News Detection in Social Networks via Crowd
  Signals}. In \bibinfo{booktitle}{\emph{Companion Proceedings of the The Web
  Conference 2018}} (Lyon, France) \emph{(\bibinfo{series}{WWW '18})}.
  \bibinfo{publisher}{International World Wide Web Conferences Steering
  Committee}, \bibinfo{address}{Republic and Canton of Geneva, CHE},
  \bibinfo{pages}{517–524}.
\newblock
\showISBNx{9781450356404}
\urldef\tempurl%
\url{https://doi.org/10.1145/3184558.3188722}
\showDOI{\tempurl}


\bibitem[\protect\citeauthoryear{Vlachos and Riedel}{Vlachos and
  Riedel}{2014}]%
        {Vlachos2014}
\bibfield{author}{\bibinfo{person}{Andreas Vlachos} {and}
  \bibinfo{person}{Sebastian Riedel}.} \bibinfo{year}{2014}\natexlab{}.
\newblock \showarticletitle{Fact Checking: Task definition and dataset
  construction}. In \bibinfo{booktitle}{\emph{Proceedings of the {ACL} 2014
  Workshop on Language Technologies and Computational Social Science}}.
  \bibinfo{publisher}{Association for Computational Linguistics},
  \bibinfo{address}{Baltimore, MD, USA}, \bibinfo{pages}{18--22}.
\newblock
\urldef\tempurl%
\url{https://doi.org/10.3115/v1/W14-2508}
\showDOI{\tempurl}


\bibitem[\protect\citeauthoryear{Vosoughi, Roy, and Aral}{Vosoughi
  et~al\mbox{.}}{2018}]%
        {Vosoughi1146}
\bibfield{author}{\bibinfo{person}{Soroush Vosoughi}, \bibinfo{person}{Deb
  Roy}, {and} \bibinfo{person}{Sinan Aral}.} \bibinfo{year}{2018}\natexlab{}.
\newblock \showarticletitle{The spread of true and false news online}.
\newblock \bibinfo{journal}{\emph{Science}} \bibinfo{volume}{359},
  \bibinfo{number}{6380} (\bibinfo{year}{2018}), \bibinfo{pages}{1146--1151}.
\newblock
\showISSN{0036-8075}
\urldef\tempurl%
\url{https://doi.org/10.1126/science.aap9559}
\showDOI{\tempurl}


\bibitem[\protect\citeauthoryear{Wang, Cao, de~Melo, and Liu}{Wang
  et~al\mbox{.}}{2016}]%
        {wang-etal-2016-relation}
\bibfield{author}{\bibinfo{person}{Linlin Wang}, \bibinfo{person}{Zhu Cao},
  \bibinfo{person}{Gerard de Melo}, {and} \bibinfo{person}{Zhiyuan Liu}.}
  \bibinfo{year}{2016}\natexlab{}.
\newblock \showarticletitle{Relation Classification via Multi-Level Attention
  {CNN}s}. In \bibinfo{booktitle}{\emph{Proceedings of the 54th Annual Meeting
  of the Association for Computational Linguistics (Volume 1: Long Papers)}}.
  \bibinfo{publisher}{Association for Computational Linguistics},
  \bibinfo{address}{Berlin, Germany}, \bibinfo{pages}{1298--1307}.
\newblock
\urldef\tempurl%
\url{https://doi.org/10.18653/v1/P16-1123}
\showDOI{\tempurl}


\bibitem[\protect\citeauthoryear{Wardle and Derakhshan}{Wardle and
  Derakhshan}{2017}]%
        {wardle2017information}
\bibfield{author}{\bibinfo{person}{Claire Wardle} {and}
  \bibinfo{person}{Hossein Derakhshan}.} \bibinfo{year}{2017}\natexlab{}.
\newblock \bibinfo{booktitle}{\emph{Information disorder: Toward an
  interdisciplinary framework for research and policy making}}.
\newblock \bibinfo{type}{{T}echnical {R}eport}. \bibinfo{institution}{Council
  of Europe Report}.
\newblock


\bibitem[\protect\citeauthoryear{Weedon, Nuland, and Stamos}{Weedon
  et~al\mbox{.}}{2017}]%
        {weedon2017information}
\bibfield{author}{\bibinfo{person}{Jen Weedon}, \bibinfo{person}{William
  Nuland}, {and} \bibinfo{person}{Alex Stamos}.}
  \bibinfo{year}{2017}\natexlab{}.
\newblock \bibinfo{booktitle}{\emph{Information Operations and Facebook}}.
\newblock \bibinfo{type}{{T}echnical {R}eport}. \bibinfo{institution}{Facebook,
  Inc.}
\newblock


\bibitem[\protect\citeauthoryear{Weikum and Theobald}{Weikum and
  Theobald}{2010}]%
        {weikum_info}
\bibfield{author}{\bibinfo{person}{Gerhard Weikum} {and}
  \bibinfo{person}{Martin Theobald}.} \bibinfo{year}{2010}\natexlab{}.
\newblock \showarticletitle{From Information to Knowledge: Harvesting Entities
  and Relationships from Web Sources}. In \bibinfo{booktitle}{\emph{Proceedings
  of the Twenty-Ninth ACM SIGMOD-SIGACT-SIGART Symposium on Principles of
  Database Systems}} (Indianapolis, Indiana, USA) \emph{(\bibinfo{series}{PODS
  '10})}. \bibinfo{publisher}{Association for Computing Machinery},
  \bibinfo{address}{New York, NY, USA}, \bibinfo{pages}{65–76}.
\newblock
\showISBNx{9781450300339}
\urldef\tempurl%
\url{https://doi.org/10.1145/1807085.1807097}
\showDOI{\tempurl}


\bibitem[\protect\citeauthoryear{Wu and He}{Wu and He}{2019}]%
        {wu2019enriching}
\bibfield{author}{\bibinfo{person}{Shanchan Wu} {and} \bibinfo{person}{Yifan
  He}.} \bibinfo{year}{2019}\natexlab{}.
\newblock \showarticletitle{Enriching pre-trained language model with entity
  information for relation classification}. In
  \bibinfo{booktitle}{\emph{Proceedings of the 28th ACM International
  Conference on Information and Knowledge Management}}.
  \bibinfo{pages}{2361--2364}.
\newblock


\bibitem[\protect\citeauthoryear{Wu, Agarwal, Li, Yang, and Yu}{Wu
  et~al\mbox{.}}{2014}]%
        {wu2014toward}
\bibfield{author}{\bibinfo{person}{You Wu}, \bibinfo{person}{Pankaj~K.
  Agarwal}, \bibinfo{person}{Chengkai Li}, \bibinfo{person}{Jun Yang}, {and}
  \bibinfo{person}{Cong Yu}.} \bibinfo{year}{2014}\natexlab{}.
\newblock \showarticletitle{Toward Computational Fact-Checking}.
\newblock \bibinfo{journal}{\emph{Proc. VLDB Endow.}} \bibinfo{volume}{7},
  \bibinfo{number}{7} (\bibinfo{date}{March} \bibinfo{year}{2014}),
  \bibinfo{pages}{589--600}.
\newblock
\showISSN{2150-8097}
\urldef\tempurl%
\url{https://doi.org/10.14778/2732286.2732295}
\showDOI{\tempurl}


\bibitem[\protect\citeauthoryear{Xiao and Liu}{Xiao and Liu}{2016}]%
        {xiao-liu-2016-semantic}
\bibfield{author}{\bibinfo{person}{Minguang Xiao} {and} \bibinfo{person}{Cong
  Liu}.} \bibinfo{year}{2016}\natexlab{}.
\newblock \showarticletitle{Semantic Relation Classification via Hierarchical
  Recurrent Neural Network with Attention}. In
  \bibinfo{booktitle}{\emph{Proceedings of {COLING} 2016, the 26th
  International Conference on Computational Linguistics: Technical Papers}}.
  \bibinfo{publisher}{The COLING 2016 Organizing Committee},
  \bibinfo{address}{Osaka, Japan}, \bibinfo{pages}{1254--1263}.
\newblock


\bibitem[\protect\citeauthoryear{Xu and Barbosa}{Xu and Barbosa}{2019}]%
        {xu-barbosa-2019-connecting}
\bibfield{author}{\bibinfo{person}{Peng Xu} {and} \bibinfo{person}{Denilson
  Barbosa}.} \bibinfo{year}{2019}\natexlab{}.
\newblock \showarticletitle{Connecting Language and Knowledge with
  Heterogeneous Representations for Neural Relation Extraction}. In
  \bibinfo{booktitle}{\emph{Proceedings of the 2019 Conference of the North
  {A}merican Chapter of the Association for Computational Linguistics: Human
  Language Technologies, Volume 1 (Long and Short Papers)}}.
  \bibinfo{publisher}{Association for Computational Linguistics},
  \bibinfo{address}{Minneapolis, Minnesota}, \bibinfo{pages}{3201--3206}.
\newblock
\urldef\tempurl%
\url{https://doi.org/10.18653/v1/N19-1323}
\showDOI{\tempurl}


\bibitem[\protect\citeauthoryear{Yamada, Shindo, Takeda, and Takefuji}{Yamada
  et~al\mbox{.}}{2016}]%
        {yamada2016joint}
\bibfield{author}{\bibinfo{person}{Ikuya Yamada}, \bibinfo{person}{Hiroyuki
  Shindo}, \bibinfo{person}{Hideaki Takeda}, {and} \bibinfo{person}{Yoshiyasu
  Takefuji}.} \bibinfo{year}{2016}\natexlab{}.
\newblock \showarticletitle{Joint Learning of the Embedding of Words and
  Entities for Named Entity Disambiguation}. In
  \bibinfo{booktitle}{\emph{Proceedings of The 20th SIGNLL Conference on
  Computational Natural Language Learning}}. \bibinfo{publisher}{Association
  for Computational Linguistics}, \bibinfo{address}{209 N. Eighth Street,
  Stroudsburg PA 18360, USA}, \bibinfo{pages}{250--259}.
\newblock


\bibitem[\protect\citeauthoryear{Yu, Gadiraju, Fetahu, Lehmberg, Ritze, and
  DIetze}{Yu et~al\mbox{.}}{2018}]%
        {knowmore_yu}
\bibfield{author}{\bibinfo{person}{Ran Yu}, \bibinfo{person}{Ujwal Gadiraju},
  \bibinfo{person}{Besnik Fetahu}, \bibinfo{person}{Oliver Lehmberg},
  \bibinfo{person}{Dominique Ritze}, {and} \bibinfo{person}{Stefan DIetze}.}
  \bibinfo{year}{2018}\natexlab{}.
\newblock \bibinfo{title}{{KnowMore - Knowledge base augmentation with
  structured web markup}}.
\newblock , \bibinfo{numpages}{159--180}~pages.
\newblock
\showISSN{22104968}
\urldef\tempurl%
\url{https://doi.org/10.3233/SW-180304}
\showDOI{\tempurl}


\bibitem[\protect\citeauthoryear{Zhang, Zhong, Chen, Angeli, and Manning}{Zhang
  et~al\mbox{.}}{2017}]%
        {zhang2017tacred}
\bibfield{author}{\bibinfo{person}{Yuhao Zhang}, \bibinfo{person}{Victor
  Zhong}, \bibinfo{person}{Danqi Chen}, \bibinfo{person}{Gabor Angeli}, {and}
  \bibinfo{person}{Christopher~D. Manning}.} \bibinfo{year}{2017}\natexlab{}.
\newblock \showarticletitle{Position-aware Attention and Supervised Data
  Improve Slot Filling}. In \bibinfo{booktitle}{\emph{Proceedings of the 2017
  Conference on Empirical Methods in Natural Language Processing}}.
  \bibinfo{publisher}{Association for Computational Linguistics},
  \bibinfo{address}{Copenhagen, Denmark}, \bibinfo{pages}{35--45}.
\newblock
\urldef\tempurl%
\url{https://doi.org/10.18653/v1/D17-1004}
\showDOI{\tempurl}


\bibitem[\protect\citeauthoryear{{Zhao, Zilong}, {Zhao, Jichang}, {Sano,
  Yukie}, {Levy, Orr}, {Takayasu, Hideki}, {Takayasu, Misako}, {Li, Daqing},
  {Wu, Junjie}, and {Havlin, Shlomo}}{{Zhao, Zilong} et~al\mbox{.}}{2020}]%
        {zhao2020fake}
\bibfield{author}{\bibinfo{person}{{Zhao, Zilong}}, \bibinfo{person}{{Zhao,
  Jichang}}, \bibinfo{person}{{Sano, Yukie}}, \bibinfo{person}{{Levy, Orr}},
  \bibinfo{person}{{Takayasu, Hideki}}, \bibinfo{person}{{Takayasu, Misako}},
  \bibinfo{person}{{Li, Daqing}}, \bibinfo{person}{{Wu, Junjie}}, {and}
  \bibinfo{person}{{Havlin, Shlomo}}.} \bibinfo{year}{2020}\natexlab{}.
\newblock \showarticletitle{Fake news propagates differently from real news
  even at early stages of spreading}.
\newblock \bibinfo{journal}{\emph{EPJ Data Sci.}} \bibinfo{volume}{9},
  \bibinfo{number}{1} (\bibinfo{year}{2020}), \bibinfo{pages}{7}.
\newblock
\urldef\tempurl%
\url{https://doi.org/10.1140/epjds/s13688-020-00224-z}
\showDOI{\tempurl}


\end{thebibliography}

\clearpage
\appendix

\section{Selected ClaimReview Claims}
\begin{table}[h!]
\caption{Selected ClaimReview claims, the relation they contain, and the relation predicted by the model. The text bold indicates the entities participating in the relation. The AUC of the relation classification task is $0.958$.}
\footnotesize
\begin{tabular}{lp{10cm}lllc}
\toprule
\textbf{ID} & \textbf{Claim} & \textbf{Actual}* & \textbf{Predicted}* & \textbf{Rating} & \textbf{Claim $\equiv$ Triple} \\
\midrule                      
1 & \textbf{Malaysian}-born Senator \textbf{Penny Wong} ineligible for Australian parliament & POB & DOB & False \\
2 & Donald Trump says \textbf{President Obama}'s grandmother in Kenya said he was born in \textbf{Kenya} and she was there and witnessed the birth. & POB & Institution & False & \checkmark\\
3 & Donald Trump says his father, \textbf{Fred Trump}, was born in a very wonderful place in \textbf{Germany}. & POB & POB & False & \checkmark\\
4 & \textbf{Barack Obama} was born in the \textbf{United States}. & POB & POB & True & \checkmark\\
\midrule
5 & \textbf{Barron Trump} was born in \textbf{March 2006} and Melania wasn't a legal citizen until July 2006. So under this executive order, his own son wouldn't be an American citizen. & DOB & POB & False\\
6 & \textbf{Isabelle Duterte} was born on \textbf{January 26, 2002}, which makes her only 15 years old today. & DOB & DOB & False\\
\midrule
7 & \textbf{Tej Pratap Yadav} receives a \textbf{doctorate degree} from Takshsila University in Bihar & education & education & False & \checkmark\\
8 & \textbf{ Smriti Irani} has a \textbf{MA degree}. & education & institution & False & \checkmark \\
9 & \textbf{Melania Trump} lied under oath in 2013 about graduating from college with a \textbf{bachelor's degree} in architecture. &    education & institution & False  \\
10 & Did \textbf{Michelle Obama} recently earn a \textbf{doctorate degree} in law? & education & education & False & \checkmark \\
11 & \textbf{Pravin Gordhan} does not have a \textbf{degree}. &    education & education & False & \checkmark\\
12 & \textbf{Alexandria Ocasio-Cortez}'s \textbf{economics degree} recalled. & education & institution & False & \checkmark\\
13 & Ilocos Norte Governor \textbf{Imee Marcos} claimed on January 16 that she earned a \textbf{degree} from Princeton University. &    education & education & False \\
\midrule
14 & Ilocos Norte Governor \textbf{Imee Marcos} claimed on January 16 that she earned a degree from \textbf{Princeton University}. &  institution & institution & False & \checkmark \\
15 & \textbf{Tej Pratap Yadav} receives a doctorate degree from \textbf{Takshsila University} in Bihar. & institution & education & False  \\
16 & \textbf{Patrick Murphy} embellished, according to reports, his \textbf{University of Miami} academic achievement. &  institution & institution & True  \\
17 & Mahmoud Abbas, Ali Khamenei, and \textbf{Vladimir Putin} met each other in the class of 1968 at \textbf{Patrice Lumumba University} in Moscow &  institution & institution & False \\
18 & \textbf{Mahmoud Abbas}, Ali Khamenei, and Vladimir Putin met each other in the class of 1968 at \textbf{Patrice Lumumba University} in Moscow &  institution & institution & False\\
19 & Mahmoud Abbas, \textbf{Ali Khamenei}, and Vladimir Putin met each other in the class of 1968 at \textbf{Patrice Lumumba University} in Moscow &  institution & institution & False \\
20 & \textbf{Maria Butina} is a human rights activist, a student of the \textbf{American University}, and the most relevant is that she is a person who did not work (collaborate) with the Russian state bodies. &  institution & institution & False\\
21 & Ilocos Norte Governor \textbf{Imee Marcos} graduated cum laude from the \textbf{University of the Philippines} (UP) College of Law. &  institution & institution & False \\
22 & \textbf{David Hogg} graduated from \textbf{Redondo Shores High School} in 2015. &  institution & institution & False & \checkmark \\
\midrule
23 & Sadhvi Pragya Singh Thakur said \textbf{Manohar Parrikar} died of cancer because he allowed the consumption of beef in \textbf{Goa}. & POD & POD & False \\
24 & Fox star \textbf{Tucker Carlson} in critical condition (then died) after head on collision driving home in \textbf{Washington D.C.} & POD & POD & False & \checkmark \\
25 & \textbf{Nasser Al Kharafi} died in \textbf{Kuwait}. & POD & POD & False & \checkmark\\
26 & DCP \textbf{Amit Sharma} passed away in \textbf{Delhi} riots & POD & institution & False & \checkmark\\
27 & It is being claimed that \textbf{Jason Statham} was murdered at his home in \textbf{New York} by assailants who broke into his mansion. & POD & POD & False\\
28 & Actor \textbf{Robert Downey Jr.} died in a car crash stunt in \textbf{Hollywood} on July 8. & POD & POD & False\\
\bottomrule
\multicolumn{6}{l}{
* DOB = Date of Birth, POB = Place of Birth, POD = Place of Death}
\label{table:claimreview}
\end{tabular}
\end{table}

\end{document}